# Coexistence of Weak Ferromagnetism and Ferroelectricity in the High Pressure LiNbO$_3$-type Phase of FeTiO$_3$

T. Varga[1], A. Kumar[2], E. Vlahos[2], S. Denev,[2] M. Park[3], S. Hong[1], T. Sanehira[4], Y. Wang,[4] C. J. Fennie,[5] S. K. Streiffer,[6] X. Ke[7], P. Schiffer,[7] V. Gopalan,[2] and J.F. Mitchell[1]

[1]Materials Science Division, Argonne National Laboratory, Argonne IL 60439; [2]Department of Materials Science and Engineering, Pennsylvania State University, University Park, PA 16802; [3]Department of Materials Science and Engineering, Korea Advanced Institute of Science and Technology, Daejon 305-701, Korea. [4]Center for Advanced Radiation Sources, The University of Chicago, Chicago, IL 60637; [5]Department of Applied and Engineering Physics, Cornell University, Ithaca, NY 14853; [6]Center for Nanoscale Materials, Argonne National Laboratory, Argonne IL 60439; [7]Department of Physics and Materials Research Institute, Pennsylvania State University, University Park, PA 16802

We report the magnetic and electrical characteristics of a polycrystalline specimen of FeTiO$_3$ synthesized at high pressure that is isostructural with acentric LiNbO$_3$ (LBO). Magnetometry, piezoresponse force microscopy, and optical second harmonic generation demonstrate that FeTiO$_3$-II is ferroelectric at and below room temperature and weakly ferromagnetic below ~120 K. These results validate symmetry-based materials design criteria and first principles calculations of coexistence between ferroelectricity and weak ferromagnetism in a series of transition metal titanates crystallizing in the LBO structure. The high-pressure form of FeTiO$_3$ stands out as a rare example of a ferroelectric exhibiting weak ferromagnetism generated by a Dzyaloshinskii-Moriya interaction.

Multiferroics are materials in which seemingly contra-indicated ferroic properties, e.g., magnetism and polar order, coexist.[1, 2] Magnetic ferroelectrics for which the different ferroic orders couple, either macroscopically through interfacial magnetostriction [3, 4] or microscopically via exchange striction,[5] may be promising materials for applications in memories, sensors, actuators, and other multifunctional devices. They also offer a rich opportunity to study fundamental aspects of spin-lattice coupling. In the case of bulk materials, several neutron diffraction studies point to a spiral magnetic state as an essential ingredient for coupling between magnetism and ferroelectricity.[5-7] Phenomenological [8] as well as first principles explanations [9] of the connection between ferroelectricity and the magnetic spiral link polar and magnetic orders through the product $\mathbf{P} \sim \mathbf{e} \times \mathbf{Q}$, where $\mathbf{e}$ is a unit vector along the spin rotation axis and $\mathbf{Q}$ is the spiral propagation vector. The microscopic origin of this form can be traced to the antisymmetric Dzyaloshinskii-Moriya (D-M) interaction,[1] which can lead to inhomogeneous states, such as the magnetic spiral found in BiFeO$_3$. Due to the nature of the magnetic spiral, no net ferromagnetic moment is found in such systems.

Rather than an inhomogeneous spiral state, the D-M interaction can alternatively lead to weak ferromagnetism (WFM), as observed in manganites [10] and other transition metal oxides, such as rare-earth orthoferrites.[11, 12] WFM in a material that is simultaneously ferroelectric is particularly interesting as it has been recently discussed as the best route to achieve electric field control of 180° switching of ferromagnetic domains,[13, 14] yet identifying a material with the required coupling, even in principle, has proven challenging. Recently, Fennie has argued from symmetry principles that polar order will induce a non-zero staggered D-M interaction, and hence weak ferromagnetism, when an invariant of the form $E \sim \mathbf{P} \cdot (\mathbf{L} \times \mathbf{M})$—where $\mathbf{P}$, $\mathbf{L}$, and $\mathbf{M}$ are polar, antiferromagnetic and magnetization vectors, respectively—exists in the phenomenological free energy functional of the putative high temperature antiferromagnetic, paraelectric parent. Using developed materials design criteria to guide the search for an experimental realization of this model, Fennie argued that materials crystallizing in the high pressure form, i.e., the LiNbO$_3$ phase, of FeTiO$_3$, MnTiO$_3$, and NiTiO$_3$ are candidate materials that exhibit the required coupling.[14] Additionally, first principles calculations on these materials indicate that they would have extremely high polarization, comparable to that of BiFeO$_3$,[2] making them attractive targets in the search for new multiferroic systems.

In this Letter we report the synthesis and characterization of the high pressure form of FeTiO$_3$ (FeTiO$_3$-II), which is found to be ferroelectric at and below room temperature and weakly ferromagnetic below ~120 K. These results validate Fennie's principles of microscopic materials design that predict the coexistence of weak ferromagnetism and ferroelectric polarization in this class of materials.[14] From a fundamental standpoint, this is particularly important, as FeTiO$_3$-II joins an extremely small class of ferroelectrics with WFM arising from the D-M interaction. Our results furthermore provide a significant step toward establishing FeTiO$_3$-II as a prototype bulk multiferroic whose magnetic structure can in principle be switched by reversing an applied electric field.

The high pressure, high temperature phase diagram of FeTiO$_3$ has been discussed previously, with FeTiO$_3$ in the LiNbO$_3$ structure believed to be a metastable compound.[15] The high pressure phase, FeTiO$_3$-II, was prepared using a multianvil press at the 13-ID-D beamline (GSECARS) of the Advanced Photon Source (APS). A platinum capsule was loaded with ilmenite prepared by standard ceramic techniques and pressed to 18 GPa followed by resistive heating to 1200 °C for 1 hour. Topographic imaging of the product by atomic force microcopy (AFM) reveals a typical grain size of about 400 nm (see Fig. 2a), and EDS analysis of several grains confirms the Fe:Ti stoichiometry. Synchrotron powder x-ray diffraction (SXRD) data were collected at the 11-BM-B beamline at the APS. The SXRD data were refined with the Rietveld method using the published crystal structure of FeTiO$_3$-II [16] as a starting point. Details of the crystallographic refinement can be found in the Supplementary Material. The coordination of the Ti atoms by oxygen is highly asymmetric, with three bonds at 1.849(3) Å and three at 2.121(3) Å, similar to that reported for FeTiO$_3$ by Leinenweber et al.[16] The iron coordination is considerably less distorted (d$_{Fe-O}$ = 2.175(3) Å and 2.099(3) Å). In the high pressure phase, both Fe$^{2+}$ and Ti$^{4+}$ ions move in an opposite sense along $c$ from their respective octahedral centers. This geometry can be thought of as resulting from an A$_{2u}$ distortion coordinate that takes an hypothetical R-3c structure to the observed R3c structure.

DC SQUID magnetization data measured in 1 kOe (see Supplementary Material) yield a linear $\chi^{-1}$ vs. T for 150 K < T < 300 K. The extracted p$_{eff}$ = 5.6 $\mu_B$/Fe is consistent with that reported for ambient-pressure ilmenite (p$_{eff}$ = 5.62 $\mu_B$/Fe [17]). A measured $\theta_W$ = -248 K agrees well with the first-principles predicted value of -305 K.[18] We note that $\theta_W$ and T$_N$ for ilmenite are 23 K and 55 K, respectively,[19] demonstrating a substantially modified magnetic exchange among Fe$^{2+}$ ions in the high pressure phase. $\chi^{-1}$ vs. T deviates from linearity below T=120 K signaling the onset of cooperative magnetism.

Figure 1b compares the AC magnetic susceptibility of an ilmenite sample to that of FeTiO$_3$-II. The known antiferromagnetic transition of ilmenite at ~55 K is replaced by a sharp cusp at T ~ 110 K. We also find a clear anomaly in the heat capacity (Fig. 1a) near this temperature. This thermodynamic signature—combined with a lack of frequency dependence in the 1-10 kHz range—demonstrate that below 110 K FeTiO$_3$-II is a long-range ordered antiferromagnet, in agreement with the theoretical prediction.[14] Figure 1c shows isothermal magnetization measured at 105 K, just below T$_N$. A clear hysteresis is observed in the data, indicative of a WFM component superimposed on the antiferromagnetic background. We note that $M(H)$ measured above 120 K shows no hysteresis (Fig. 1c, top left inset), demonstrating that the appearance of WFM is linked to the onset of the AFM state. A small curvature observed below ~1 kOe (Fig. 1c, top left inset) is seen in the range 5 K ≤ T ≤ 300 K and may reflect an extremely low concentration of a magnetic impurity such as Fe (found at ~1% level by XRD) or other ferrous oxides in quantities undetectable to XRD. In the bottom right inset of Fig. 1c we have subtracted the high field linear part of $M(H)$ to estimate the field dependence of the FM component alone. This analysis shows a symmetric hysteresis loop, saturating by 30 kOe at 0.008 $\mu_B$/Fe. There are two possible origins for the intrinsic WFM: (1) phase separation into discrete FM and AFM regions in the sample, as has been proposed both theoretically [20] and experimentally [21] for several doped transition metal oxides, or (2) a canting of the Fe spins away from 180°. The former scenario is unlikely based on the coincident appearance of both FM and AFM components and the lack of frequency dependence in the AC susceptibility, leading us to favor the canted state as the origin of WFM.

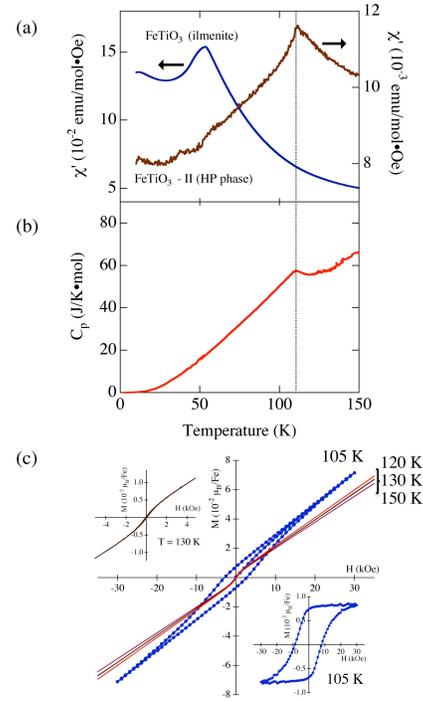

FIG 1: (a) Magnetic susceptibility of FeTiO$_3$ (ilmenite) and high-pressure FeTiO$_3$-II. (b) Specific heat of high pressure FeTiO$_3$-II. (c) Isothermal magnetization of FeTiO$_3$-II following zero-field cooling. See text for details.

We can estimate the canting angle of the WFM by invoking the symmetry arguments of Ref. [14] under the assumption that the polar domains are randomly and uniformly distributed in the polycrystalline sample.[22] According to Ref. [14] the magnetization vector, **M**, of each magnetic domain within a given polar domain will lie perpendicular to the polar vector **P**. However, in **H**=0 (or ignoring higher order single-ion anisotropy), this coupling does not fully constrain the various magnetic domains within each polar domain, as each magnetic domain need only satisfy **P**•**M** = 0 individually. Application of a sufficiently large magnetic field will rotate **M** about **P** to maximize **M**•**H** subject to this constraint, orienting **P**, **H**, and **M** coplanar within that polar domain. In this case, the projection of **M** onto **H** is $M_H = M\sin\theta$, where $\theta$ is the polar angle between **P** and **H**. Averaging over $\theta$, accounting for $T/T_C \sim 0.9$ and using the extracted estimate of $M_H = 0.008$ $\mu_B$/Fe yields $M_S = 0.04$ $\mu_B$/Fe, in excellent agreement with the value 0.03 $\mu_B$/Fe calculated in Ref. [14]. The canting angle calculated using this value of $M_S$ is 1°.

Several facts argue that the weak ferromagnetic signal is intrinsic to the FeTiO$_3$-II phase (See Supplementary Material): (1) None of the iron-containing impurity phases (all in ~1% level) found in Rietveld analysis of the the x-ray data have ferromagnetic transition temperatures near 110 K. (2) The specific heat (Fig. 1a) shows a well-defined anomaly centered at 110 K with amplitude exceeding what could be accounted for by a ~1% ferromagnetic impurity such as Fe$_3$O$_4$ (with its Verwey transition at ~120 K), and an impurity ferromagnetic phase large enough to account for our heat capacity signal would yield an orders of magnitude larger magnetic signal. (3) Measurement of a second sample prepared under different conditions yields both qualitatively and quantitatively consistent magnetic behavior.

We have explored the polar properties of FeTiO$_3$-II using piezoresponse force microscopy (PFM) and optical second harmonic generation (SHG): both techniques indicate that FeTiO$_3$-II is ferroelectric at room temperature. Out-of-plane PFM imaging has been used to confirm the detailed ferroelectric domain configuration at the nanoscale (For detailed imaging conditions, see Ref. [23]). The amplitude images in Fig. 2(b) show the varying degrees of alignment of polarization vectors to the surface normal in each domain, and reveal the various distributions of polarization vectors. The phase images (Fig. 2c) sample the direction of polarization vectors showing the expected mixture of up (dark contrast) and down (bright contrast) polar domains. This is analogous to similar domain structures of bulk polycrystalline Pb(Zr,Ti)O$_3$.[24]

A stationary-tip piezoresponse hysteresis loop obtained by applying 0.5 V$_{rms}$ to the tip while sweeping the dc voltage from -10 V to 10 V to the bottom electrode with frequency of 41.7 mHz is shown in Fig 2d. Although strong imprint of the loop, which could be due to the strong bias field created by unswitched polarization beneath the grain of interest, is apparent along both the electric field and piezoelectric coefficient axes, this measurement is sufficient to demonstrate a reversible polar response and therefore ferroelectricity. It is important to note that spatial variation of electromechanical properties was observed. This indicates that these samples are not fully homogeneous, which is not unexpected for this stage of synthesis development.

Ferroelectricity is further established in FeTiO$_3$-II via optical Second Harmonic Generation (SHG), which involves the conversion of light at a frequency $\omega$ (electric field $E^\omega$) into an optical signal at a frequency $2\omega$ by a nonlinear medium through the creation of a nonlinear polarization $P_i^{2\omega} = d_{ijk}E_j^\omega E_k^\omega$, where $d_{ijk}$ represents the nonlinear optical tensor coefficients (not to be confused with the piezoelectric tensor). SHG occurs only in the absence of inversion symmetry, which is also a necessary condition for a polar medium such as a ferroelectric. Optical SHG mapping was performed with a fundamental wave generated from a tunable Ti-sapphire laser with 65-fs pulses of wavelength 800 nm incident normal to the sample surface. 2-D mapping of the signal was done using a WITec Alpha 300 S confocal microscope.

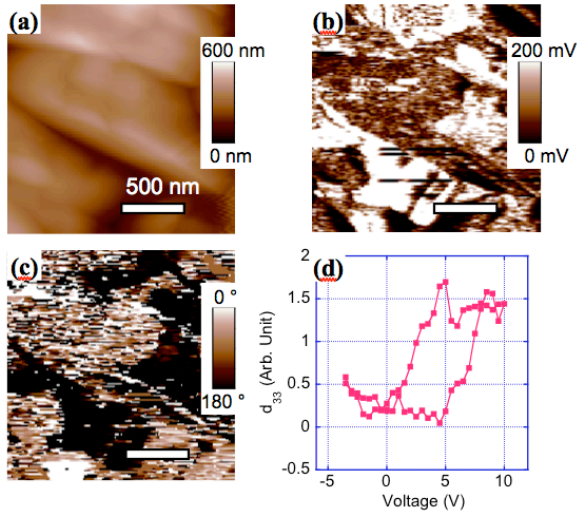

FIG 2: Surface topography (a), out-of-plane PFM amplitude (b) and phase (c) images in FeTiO$_3$ bulk crystal sample. Bright contrasts in amplitude correspond to polarization vectors strongly aligned to the surface normal (either positive or negative normal direction). Bright contrasts in phase images correspond to the down polarization whereas dark ones to the up polarization. (d) Local piezoresponse hysteresis loops were collected inside a grain.

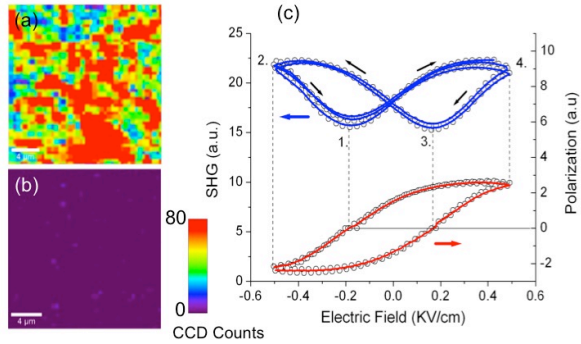

Figure 3 SHG mapping of polar regions at 296K in (a) stabilized high pressure phase of FeTiO$_3$ with LiNbO$_3$ crystal structure (non-centrosymmetric) (b) pure ilmenite phase (centrosymmetric) . (c) SHG "butterfly" hysteresis loops (blue) at a fixed point in (a) with applied electric fields at 296 K and the corresponding polarization hysteresis loop (red) (see text). The polarization of fundamental light was horizontal, and all the output SHG polarizations were detected without an exit analyzer.

While the stabilized LiNbO$_3$ phase of FeTiO$_3$ shows a strong SHG contrast (Fig. 3a), no signal was observable in the ilmenite phase (Fig. 3b). These results confirm that the high-pressure phase is polar while the ilmenite phase is nonpolar. The spatial variation of the SHG signal in Fig. 3(a) can arise from differently oriented polycrystallites as well as multi-domain structure. The SHG hysteresis loops at different spots on the sample were measured using electrodes applied on opposite edges of the sample while probing the top surface. A representative measurement is shown in Figure 3(c), and has the "butterfly" shape characteristic of the response of a ferroelectric. We reasonably exclude effects such as electric-field induced SHG (EFISH) as insignificant, since no such effects are seen in the compositionally similar Ilmenite phase under an electric field. The corresponding polarization hysteresis loop shown in Figure 3(c) can be derived from the SHG intensity vs. electric field data as follows: the SHG $I^{2\omega} \propto d_{ijk}^2 \propto (\chi_{ijks} P_s)^2$, where $\chi_{ijks}$ represents the fourth order nonlinear optical susceptibility tensor in the paraelectric phase. Though points 1 and 3 in Figure 3(c) of the butterfly loop correspond to the field axis crossings of the polarization hysteresis loops, the SHG intensity is not exactly zero at these minima, due to an incomplete cross-cancellation of the SHG intensity between antiparallel domains in the area of the sample being probed. Thus, in going from the SHG intensity to the switchable polarization loops, we first subtract a baseline intensity corresponding to a linear extrapolation between two minima (1 and 3), followed by taking the square-root of the intensity, and finally switching the sign of the result only for the segment of the butterfly loop 1-2-3 in Figure 3(c). This yields a polarization hysteresis loop that is proportional to the net *switchable* polarization within the probe region, and clearly confirms the presence of ferroelectricity. The SHG intensity was observed down to 5 K, indicating that the FeTiO$_3$-II is polar and multiferroic below the Néel temperature.

In summary, we have prepared the LiNbO$_3$ polymorph of ilmenite, FeTiO$_3$-II, at high pressure and found that it is both ferroelectric and weakly ferromagnetic. The measured magnetic transition temperatures and WFM moment are in excellent agreement with the first principles calculations.[14] These results validate the calculations of Ref. [14] and furthermore offer a strong indication that the predicted magneto-electric coupling is operable in this compound. For definitive proof of this effect, it remains to demonstrate explicitly using aligned single crystals that the magnetic and polar domains are coupled.

Work at Argonne and use of the Advanced Photon Source and the Center for Nanoscale Materials is supported by the U.S. Department of Energy Office of Science under Contract No. DE-AC02-06CH11357. Portions of this work were performed at the GEeoSoilEnviroCARS, which is supported by the National Science Foundation – Earth Sciences (EAR-0622171) and the Department of Energy – Geosciences (DE-FG01-94ER14466). PS and VG acknowledge support from the National Science Foundation grant numbers DMR-0820404, DMR-0507146, and DMR-0512165. CJF acknowledges support from the Cornell Center for Materials Research with funding from the National Science Foundation (cooperative agreement DMR 0520404).

Supplementary Material for
"**Coexistence of Weak Ferromagnetism and Polar Order in the High Pressure LiNbO$_3$-type phase of FeTiO$_3$**"

T. Varga et al.

## S1. Inverse DC Susceptibility

Fig. S1 below shows the reciprocal DC susceptibility of the primary $FeTiO_3$-II sample discussed in the manuscript. For temperatures above ~120 K, the data are linear, indicating a Curie-Weiss regime. A fit to these data yield an effective moment, $p_{eff}$, of 5.6 uM/Fe and a Weiss constant, θ, of -248 K. As discussed in the manuscript, the former is consistent with ilmenite and other $Fe^{2+}$ oxide systems, while the latter reflects strong antiferromagnetic interactions and good agreement with the theoretical value of -305 K provided in Reference 18.

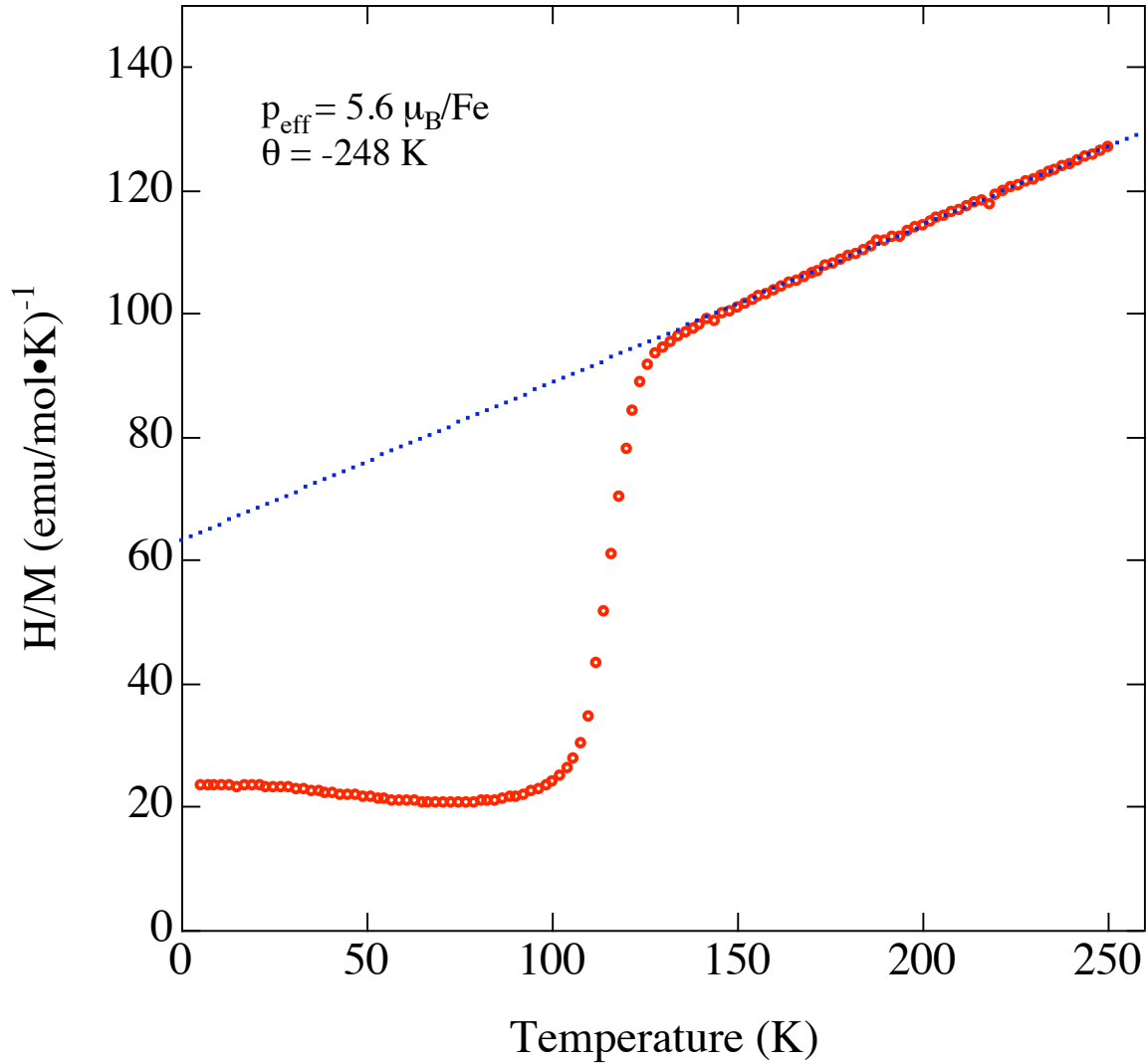

Fig. S1   Reciprocal DC magnetic susceptibility of $FeTiO_3$-II measured in 1 kOe on cooling.

## S2. Potential Impact of Magnetic Impurity Phases

As discussed in the manuscript, we identified a number of impurity phases based on analysis of the synchrotron x-ray powder diffraction data. This analysis (details are presented below in Section S5) indicated no evidence (> 1% weight fraction) for $Fe_3O_4$, magnetite, as a crystalline phase. Magnetite, which has a known Verwey transition at ~120 K, with impact on both its magnetic and electronic behavior, is of concern when discussing intrinsic versus extrinsic behavior of our $FeTiO_3$-II sample.

Despite no evidence for $Fe_3O_4$ in the diffraction data, we now consider the possibility that magnetite could be present in our sample as an amorphous phase or in very small (< 1% weight fraction), and so not observed as Bragg reflections in the x-ray data. The following discussion argues based on bulk thermodynamic and magnetic data that $Fe_3O_4$ is not responsible for the 120 K phenomenon in our sample and that this behavior is thus intrinsic to the $FeTiO_3$-II compound.

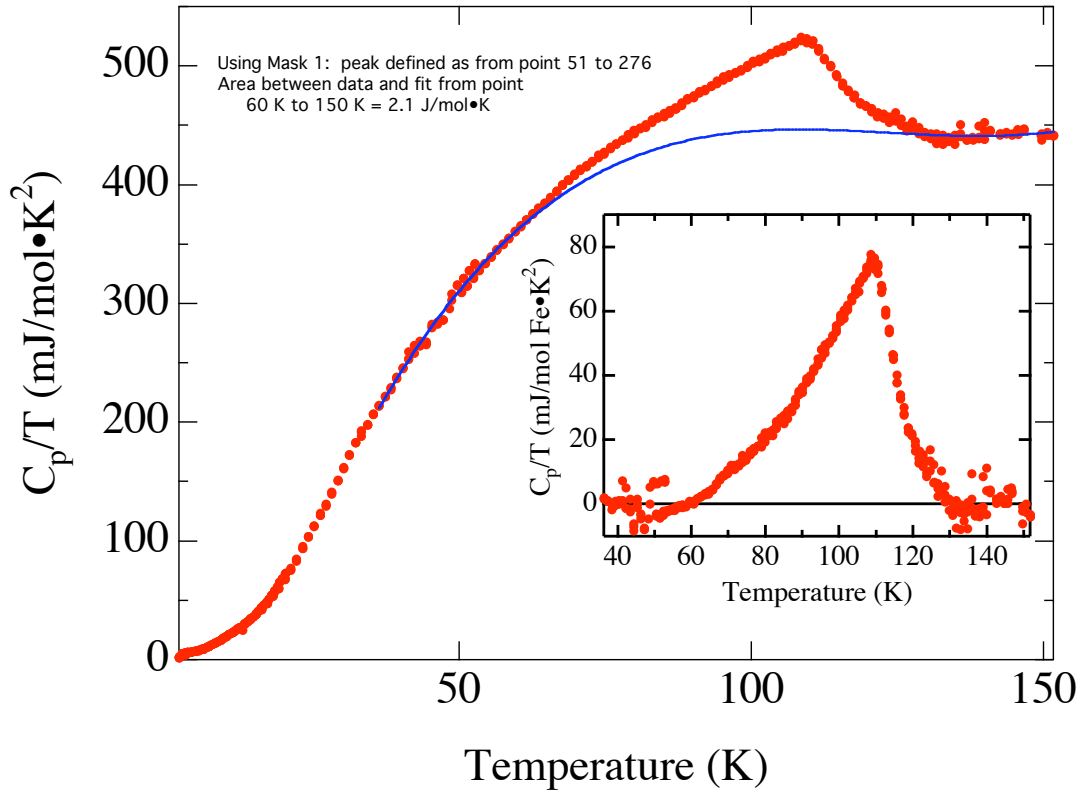

Fig. S2 Specific heat of $FeTiO_3$-II. Red points: data; blue line: phenomenological cubic polynomial background. Inset: Background-subtracted data in the range 40-150 K.

Figure S2 shows the specific heat of the primary $FeTiO_3$-II sample plotted as $C_p/T$ versus T. The blue line is a phenomenological third-order polynomial fit as 'background.' The entropy of the transition, $\Delta S_{mag}$, can be estimated as the area between the data and the background to be 2.1 J/mol•K. The specific heat of $Fe_3O_4$ has been measured by Shepherd et al. (Phys. Rev. B 31, 1107, (1985)). The entropy of the 120 K Verwey transition is reported by this group to be $\Delta S_{mag} \sim 6$ J/mole $Fe_3O_4 \cdot K$. Normalizing to a per mole Fe basis gives ~2 J/mol Fe·K. The comparable size of our measured $\Delta S_{mag}$ and that for pure $Fe_3O_4$ indicates that the transition in our data cannot be explained by contamination of $Fe_3O_4$ unless it is a *substantial majority fraction* of the sample.

However, the room temperature magnetization of our sample demonstrates that any $Fe_3O_4$ impurity must be at or below the 1% level. Shown in Fig. S3 are DC magnetization at room temperature for single crystal and thin film magnetite taken from Arora et al., *Phys. Rev. B* **77**, 134443 (2008). The saturation magnetization for the single crystal $Fe_3O_4$ at 10 kOe is ~500 emu/cm$^3$. Using the known density of $Fe_3O_4$, this corresponds to 22000 emu/mol $Fe_3O_4$ or ~ 4 $\mu_B$/f.u. = 1.33 $\mu_B$/Fe. Similar values have been reported elsewhere (Selwood, *Magnetochemistry* page 38; Trémolet de Lacheisserie, Etienne du; Gignoux, Damien; Schlenker, Michel (Eds.) *Magnetism: Materials and Applications*, p. 488)).

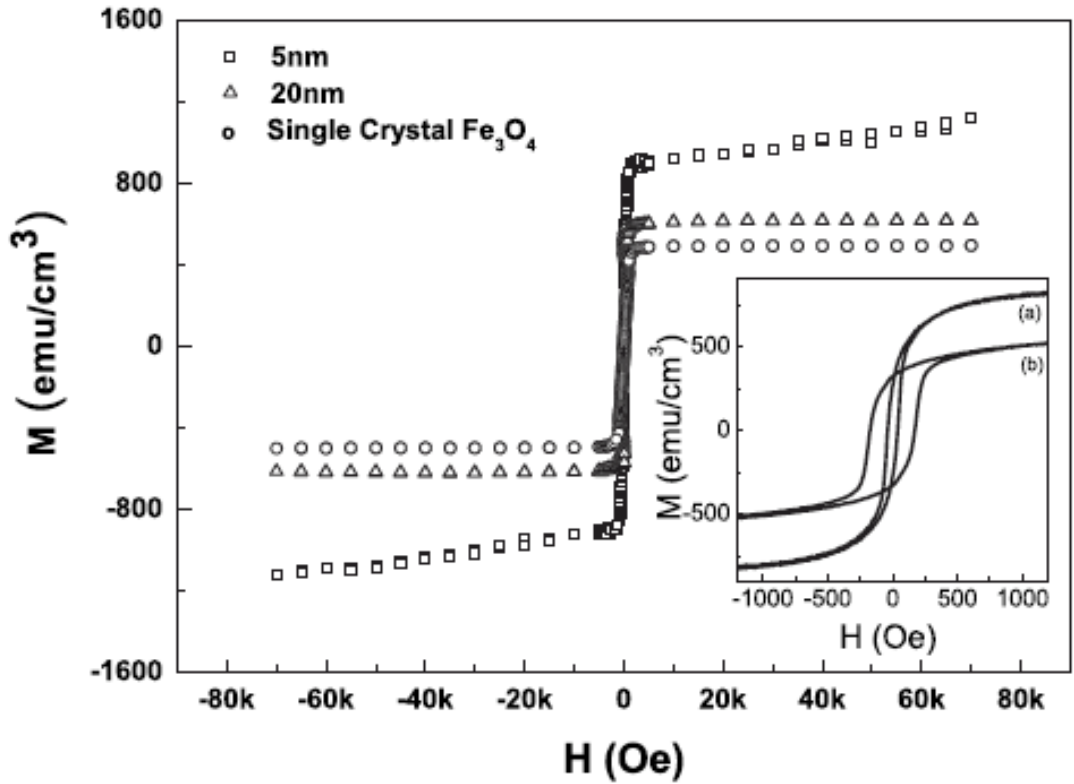

Fig. S3: DC magnetization of Fe3O4 single crystal and films. From Arora et al. *Phys. Rev. B* **77**, 134443 (2008).

In comparison, DC magnetization data measured on our $FeTiO_3$-II sample at 300 K (Fig. S4) show that our weak magnetic signal must be of order $10^{-3}$ $\mu_B$/Fe, implying < 1% $Fe_3O_4$. $Fe_3O_4$ thus does not constitute a substantial majority phase in the sample; if present at all it is in <1% abundance.

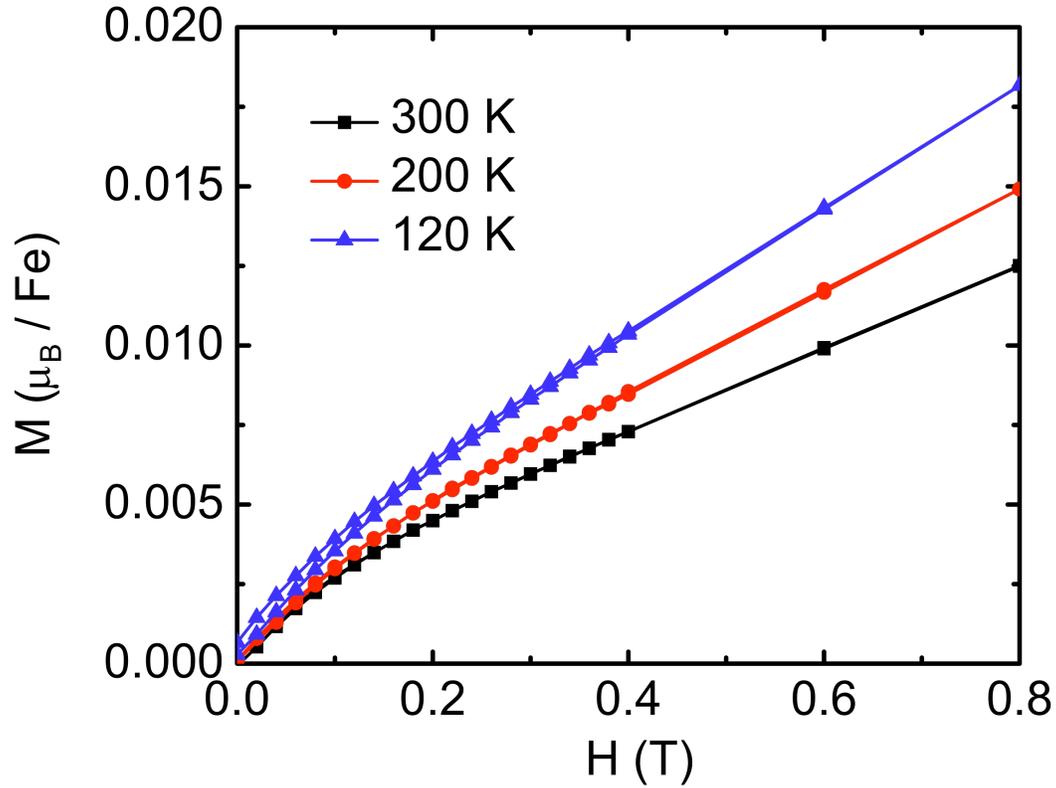

Fig. S4  DC magnetization of $FeTiO_3$-II at 300 K.  Note:  the ordinate scale is based on using the formula weight of $FeTiO_3$-II when normalizing the raw data from the magnetometer; however, the formula weights of $FeTiO_3$ and $Fe_3O_4$ differ by approximately 30%, so this cannot account for a hundred-fold or thousand-fold difference.

The assumption of an $Fe_3O_4$ impurity phase leads to contradiction between the bulk magnetization and specific heat measurements.  On this basis, we eliminate $Fe_3O_4$ as the cause of the 120 K feature in our data and argue that it is an intrinsic property of the $FeTiO_3$-II phase.

## S3. Comparison of M(T) Data for Two Samples

To ensure sample reproducibility, we synthesized a secondary sample of FeTiO$_3$-II, also at 18 GPa and 1200 $^o$C. The plots below shows M(T) measured under two fields, 10 Oe and 3 kOe on zero- and field-cooling (ZFC and FC, respectively).

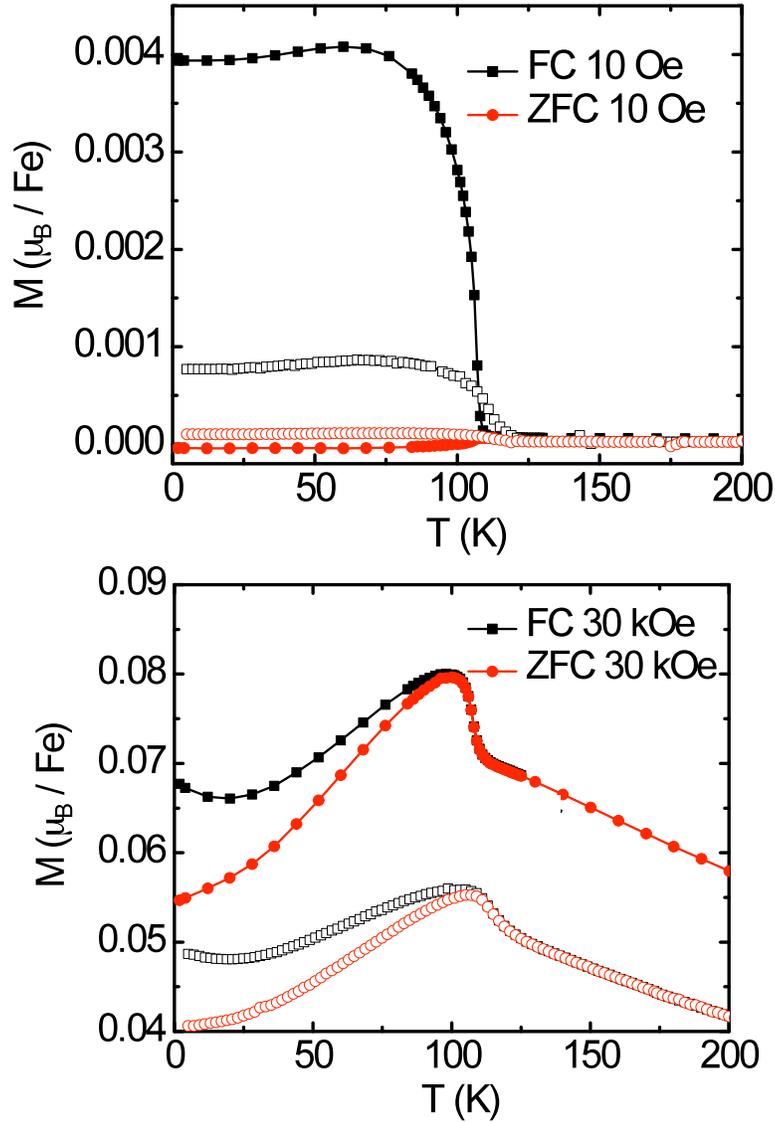

Fig. S5 DC magnetization as a function of temperature for two samples of FeTiO$_3$-II. Open symbols refer to the sample discussed in the manuscript (primary sample), while closed symbols refer to the secondary sample.

Comparison of the data indicate similar transition temperatures and qualitative temperature dependence.

## S4. Synchrotron powder Diffraction Data

To analyze the phase composition and refine the crystal structure of $FeTiO_3$-II, synchrotron powder diffraction data were collected at 11-BM-B (Advanced Photon Source) on polycrystalline sample of $FeTiO_3$-II in transmission mode. The wavelength used was 0.4581960 Å. The data shown in Fig. S6 show sharp Bragg reflections indicating that the phase(s) in the sample are well-crystallized. The flat background indicates no evidence for marked concentration of amorphous component(s).

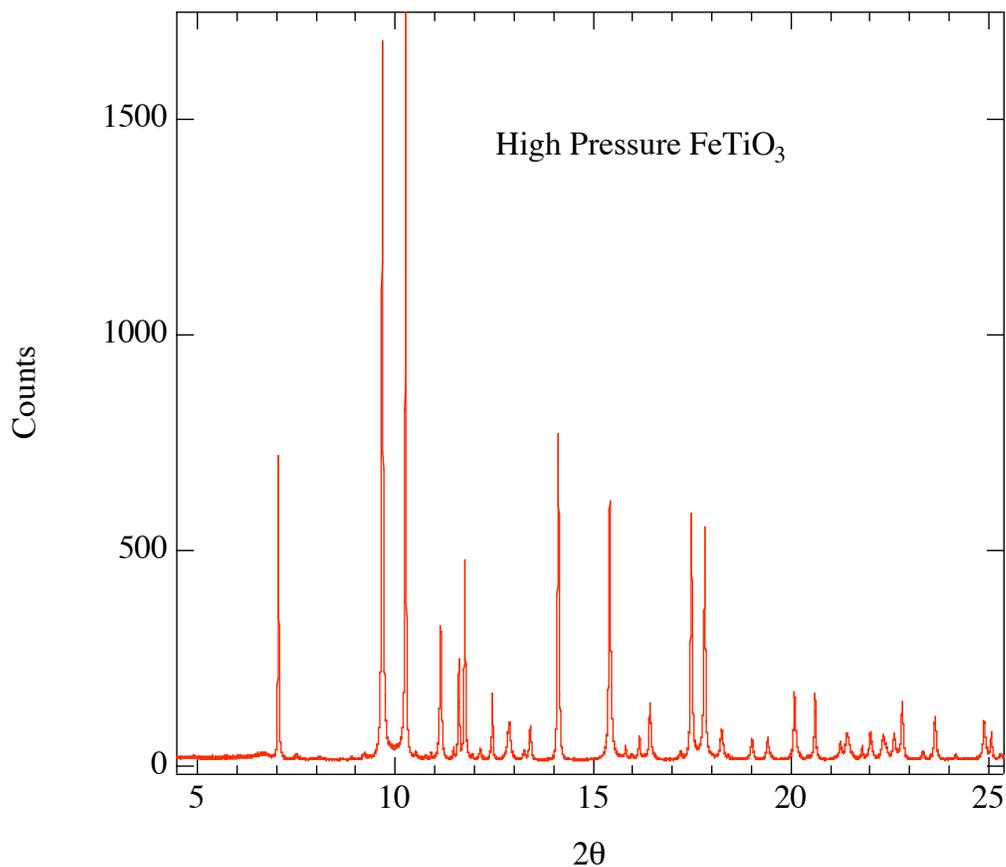

Fig. S6 Synchrotron powder diffraction data collected at the APS on the primary sample of $FeTiO_3$-II

## S5. Rietveld Refinement of Synchrotron Powder Diffraction Data

Fig. S7 shows the results of Rietveld refinement of the synchrotron powder x-ray data collected on the 11-BM-B diffractmeter at the Advanced Photon source showing all identified crystalline phases. The visual fit to the data is excellent, and refinement statistics (see Table S2 below) demonstrate the correctness of the model. Lattice constants and crystallographic parameters (Table S1) agree well with those previously published by Leinenweber et al. (*Phys. Chem. Mineral.* **22**, (1995) 251)

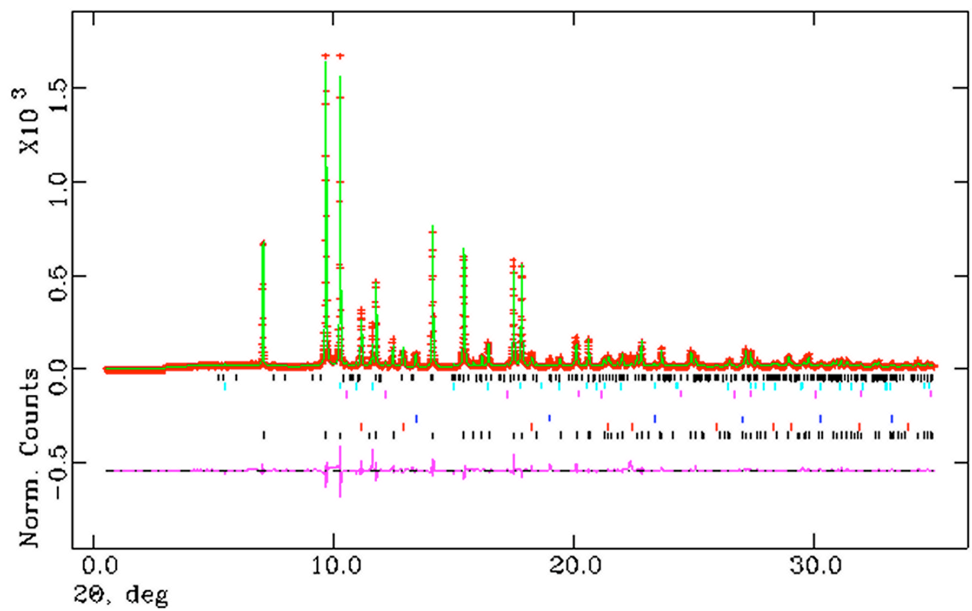

Fig. S7 Rietveld refinement of synchrotron x-ray powder diffraction data for FeTiO3-II. Red: observed data, green: calculated fit to data, purple: difference curve. Tickmarks for phases from top to bottom; black: $Fe_2TiO_5$, light blue: $Ti_2O$, purple: FeO, blue: Fe, red: Ti, black FeTiO3-II

**Table S1 :** Lattice parameters, Fe/Ti fractional occupancies, and the amounts of crystalline phases present obtained from the Rietveld fits to synchrotron powder diffraction data. Literature data for the same $FeTiO_3$ phase are also included for comparison. Space group: R3c

| Sample | a (Å) | b (Å) | c (Å) | V (Å$^3$) | Fe occup. | Ti occup. |
|---|---|---|---|---|---|---|
| This work | 5.12362(4) | 5.12362(4) | 13.7471(1) | 312.532(6) | 1.012(6) | 0.980(5) |
| Literature* | 5.12334(5) | 5.12334(5) | 13.7602(2) | 312.800(6) | 1.0 | 1.0 |

*K. Leinenweber, J. Linton, A. Navrotsky, Y. Fei, and J. B. Parise, Phys. Chem. Mineral. **22**, (1995) 251.

| Sample | FeTiO$_3$ (wt%) | Ti (wt%) | Ti$_2$O (wt%) | Fe (wt%) | FeO (wt%) | Fe$_2$TiO$_5$ (wt%) |
|---|---|---|---|---|---|---|
| This work | 88.09(10) | 6.90(7) | 3.23(6) | 1.11(2) | 0.50(22) | Trace amount |

**Table S2**. Agreement indices for the Rietveld refinement in Table S1

| | |
|---|---|
| Red. $\chi^2$ | 7.563 |
| w$R$p | 0.1463 |
| $R$(F$^2$) | 0.0601 |

The synchrotron x-ray data allow us to evaluate the potential crystalline impurity phases in the sample. We used the International Center for Diffraction Data (ICDD) PDF-4+ database to search for possible impurities based on line position and intensity matching. We then tested the reliability of candidate phases by adding them to the Rietveld refinement. This process is detailed in Figs S8 – S18. The conclusions from this analysis are that the FeTiO$_3$-II phase was contaminated by minor fractions of metallic Ti and Fe and oxides, including FeO, Ti$_2$O and Fe$_2$TiO$_5$. The weight fractions of the crystalline components are presented in Table S1.

**Ti:**
Initial phase ID showed good match to the two strongest reflections of elemental Ti (blue tickmarks).

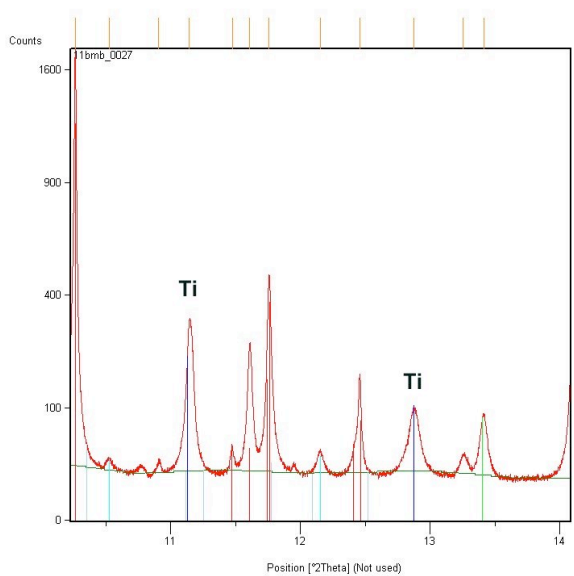

Fig. S8: ICDD database search-match revealing presence of metallic Ti.

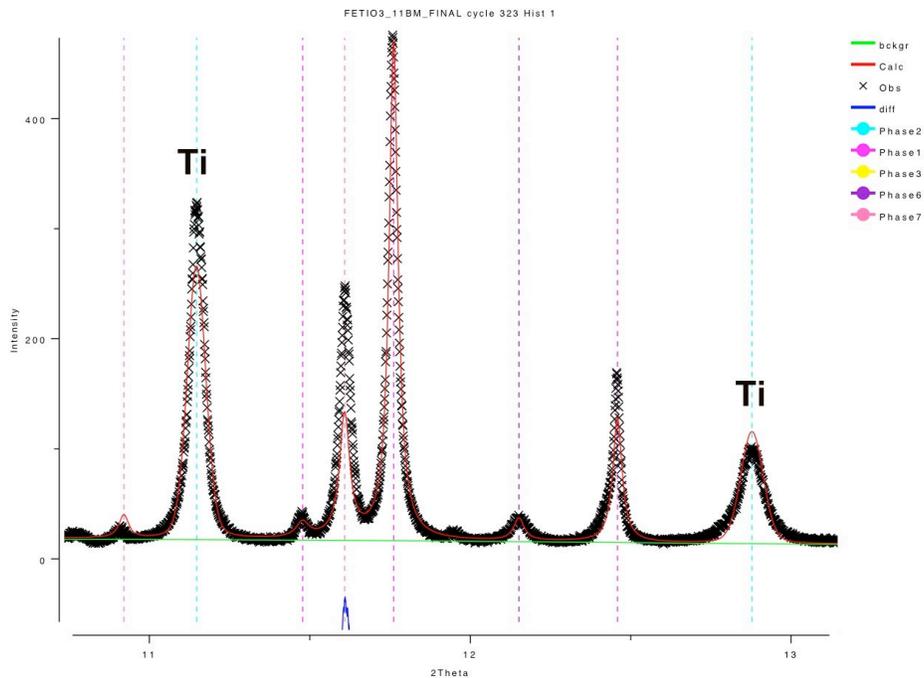

Fig. S9. Rietveld fit including Ti (refined to ~7 wt%).

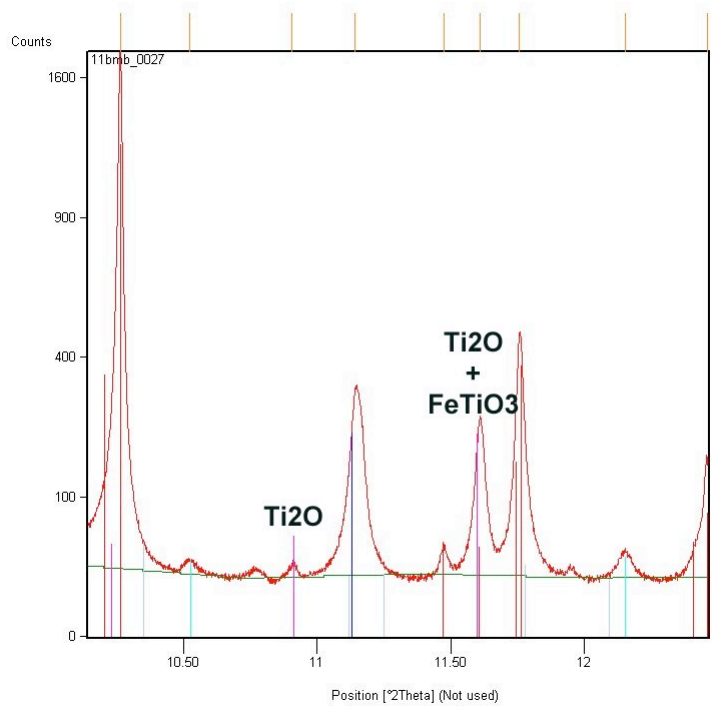

Fig. S10 ICDD database search-match revealing presence of Ti$_2$O

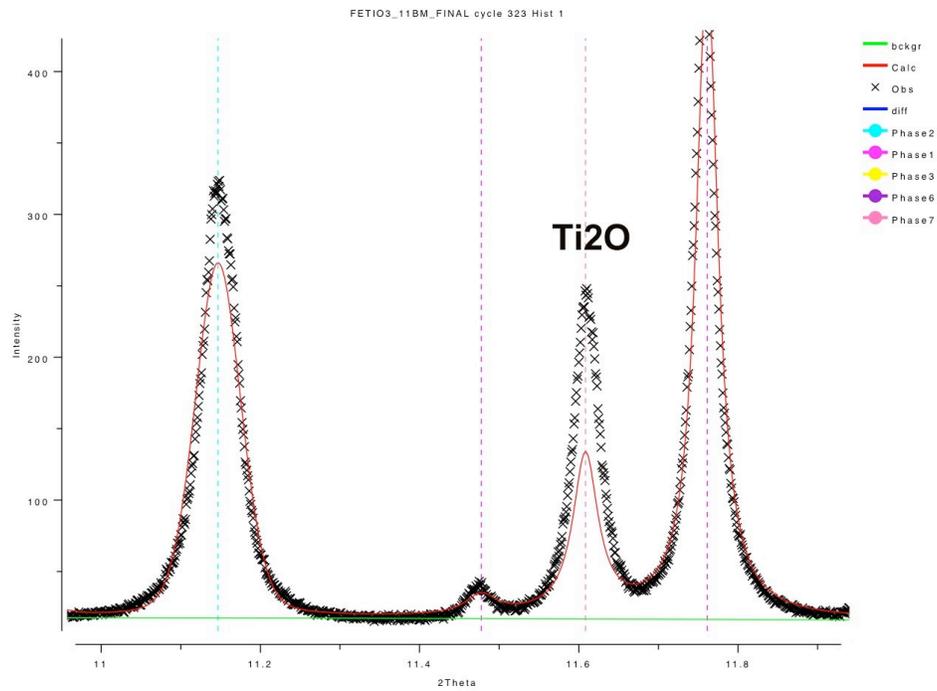

Fig. S11. Rietveld fit including $Ti_2O$ (refined to ~3 wt%).

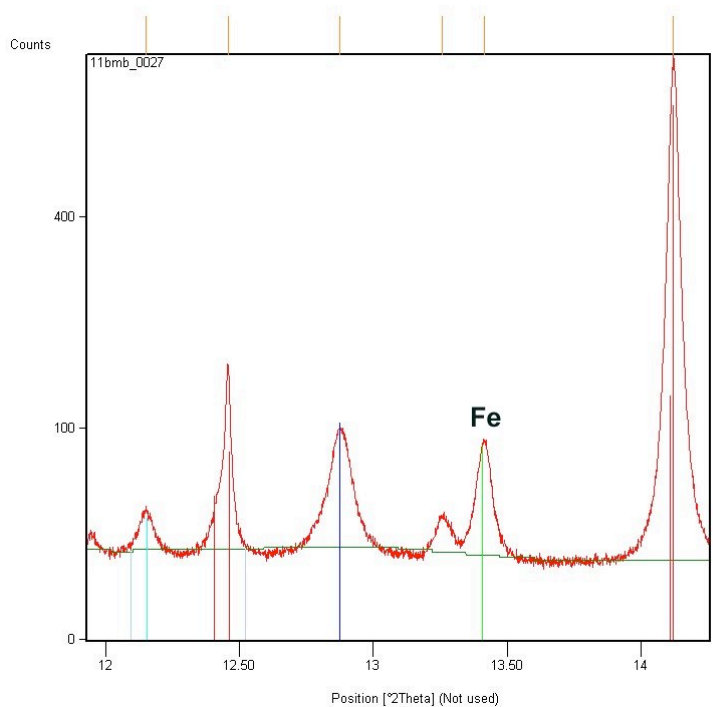

Fig. S12  ICDD database search-match revealing presence of metallic Fe.

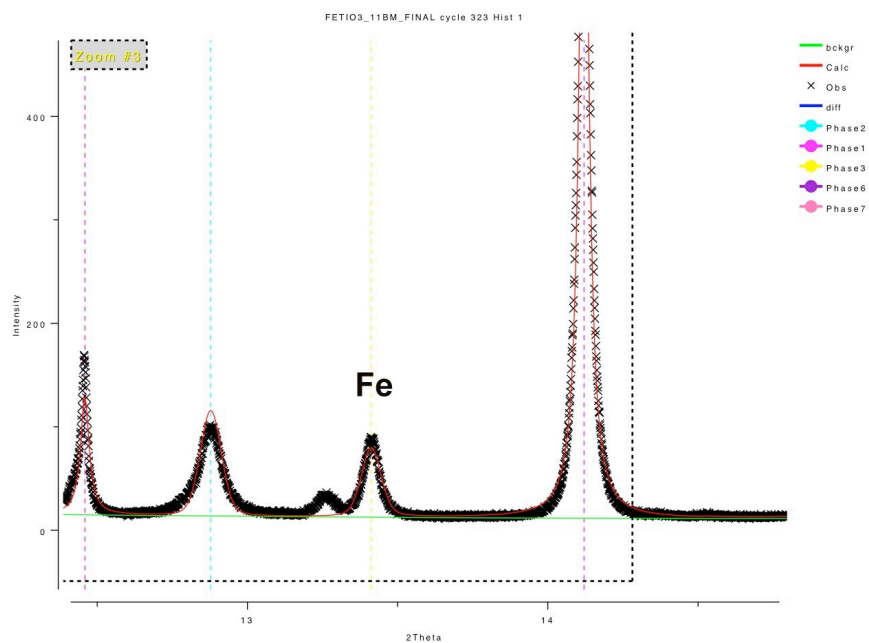

Fig. S13. Rietveld fit including Fe (refined to ~1 wt%).

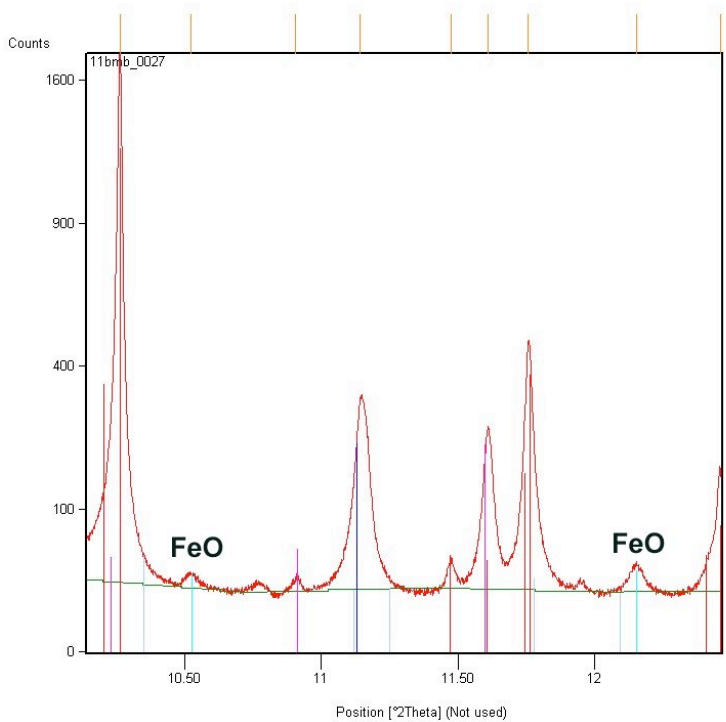

Fig. S14 ICDD database search-match revealing presence of FeO.

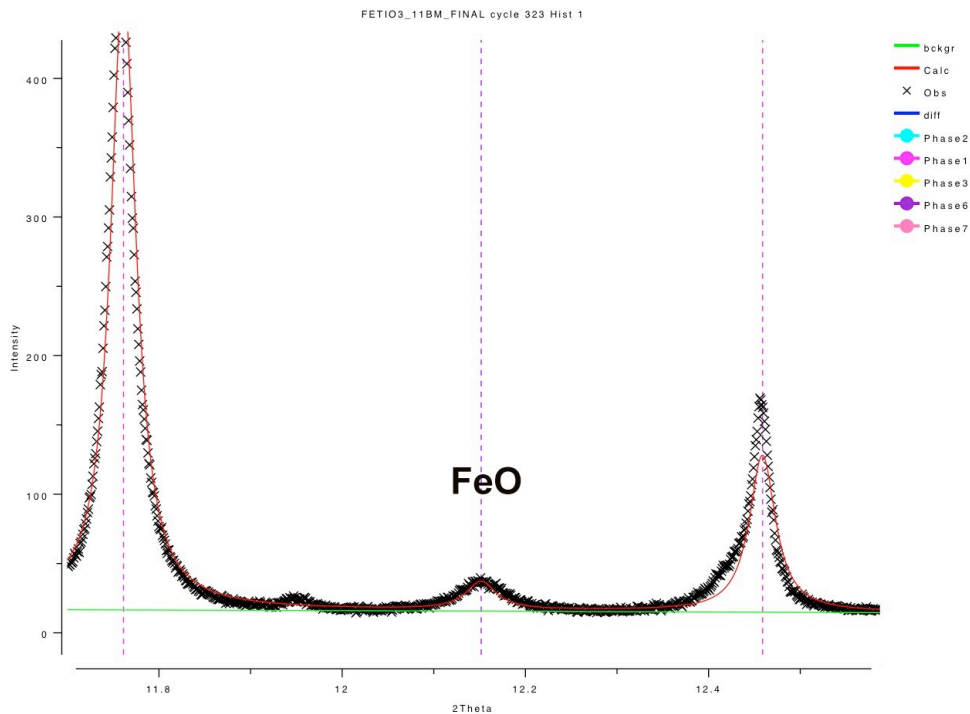

Fig. S15. Rietveld fit including FeO (refined to ~0.5 wt%).

We also checked for other likely oxides of iron that might be potential magnetic contaminants, $Fe_2O_3$ (hematite) and $Fe_3O_4$ (magnetite). The latter is particularly relevant, as it has a known magnetic and electronic transition at 120 K (Verwey transition). Both the ICDD search-match and the Rietveld fit (Fig. S16) show that while the reflections (and lattice parameters) of hematite are close to those of $FeTiO_3$, the discrepancy is clear. We note that hematite and ilmenite form a complete solid solution at ambient pressure (R.W Taylor, *Amer. Mineral.* **49**, 1016 (1964). Although the Rietveld refinement showed no evidence for Fe/Ti site mixing, we cannot rule out a very small level of such disorder. Nonetheless, we thus conclude that hematite is not present as crystalline phases above our detection level (1%). The four strongest reflections of magnetite, $Fe_3O_4$, are compared to the diffraction pattern in Fig. S17. The Rietveld fit (Fig. S18) shows that the strongest reflection ($2\theta = 10.38°$) and another weaker reflection are not matched at all. These data indicate that magnetite is not present as a crystalline phase above the 1% level.

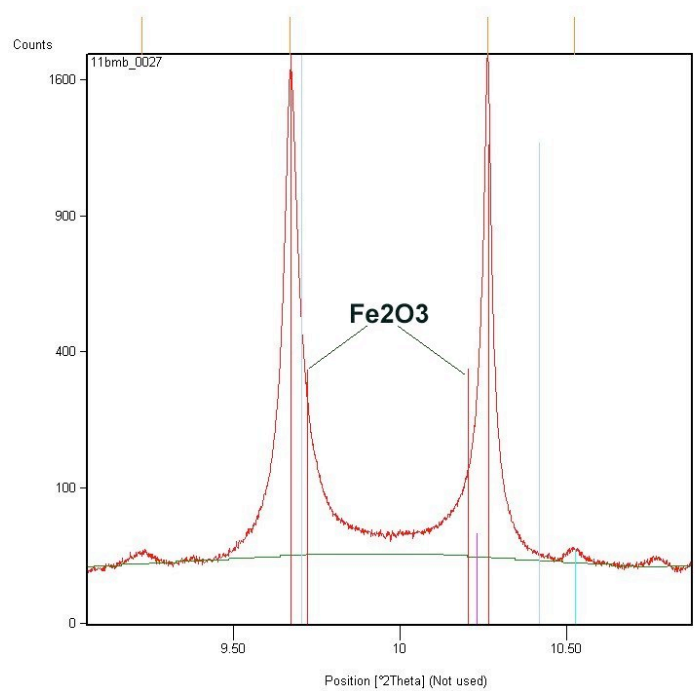

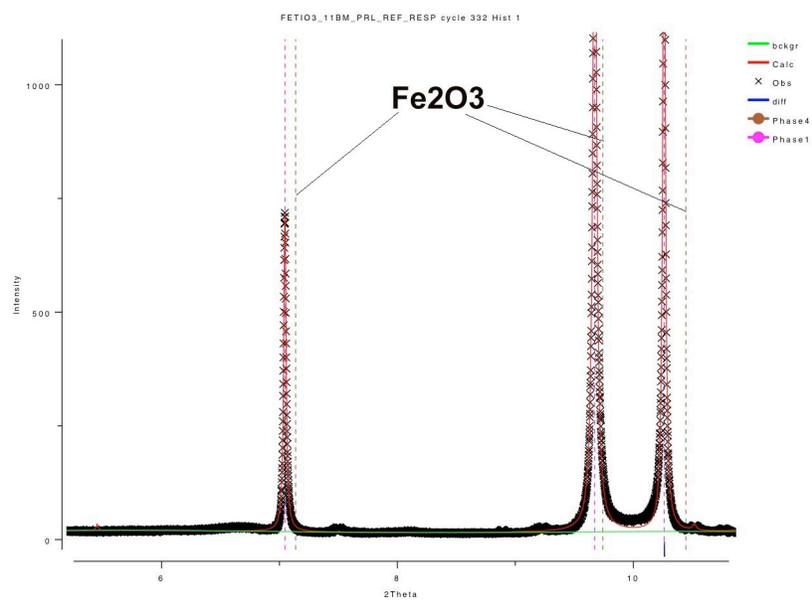

Fig S16  ICDD database search-match for $Fe_2O_3$ (top) and Rietveld refinement including for $Fe_2O_3$ as a candidate impurity phase (bottom).

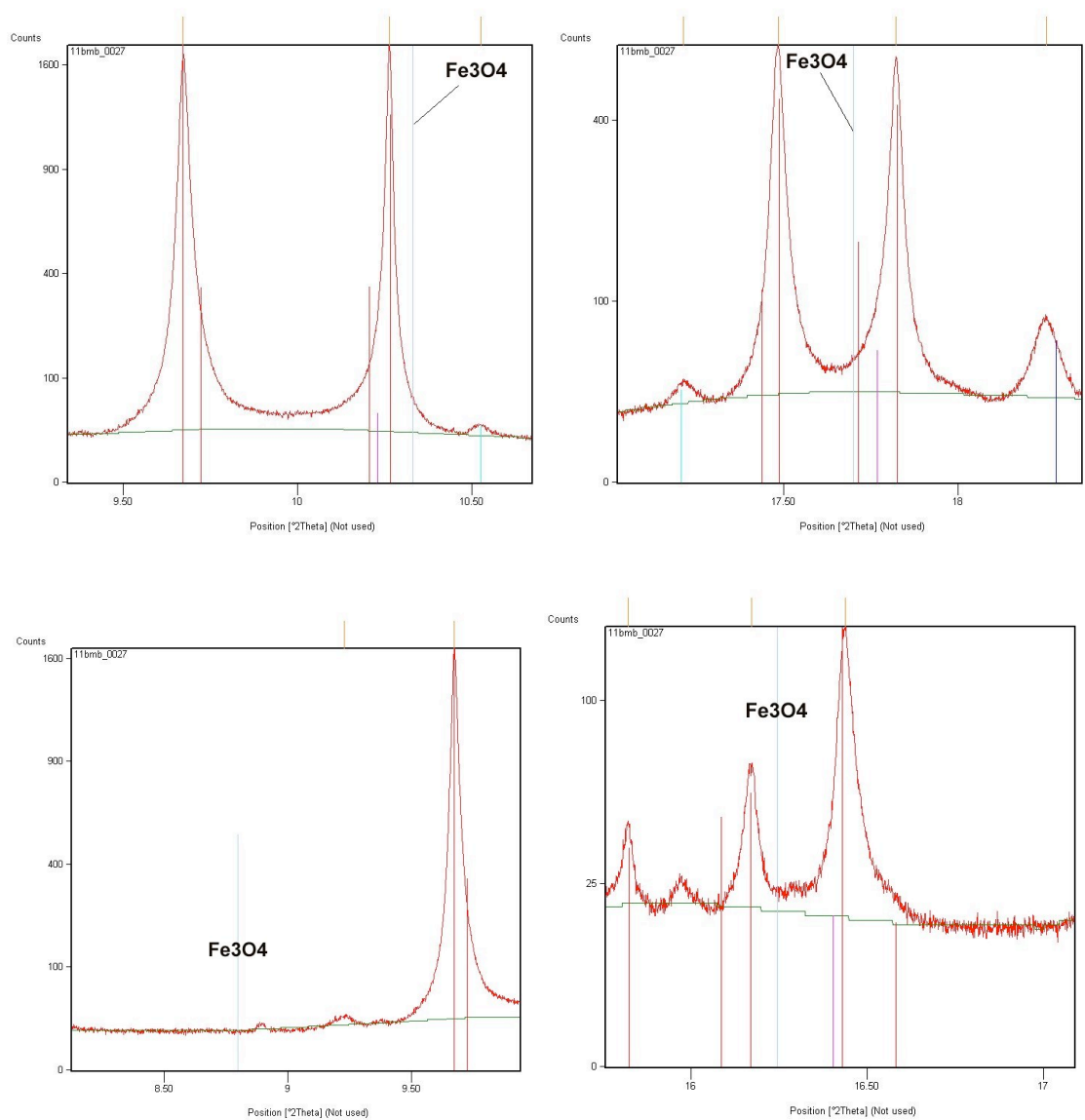

Fig. S17. ICDD search-match for $Fe_3O_4$.

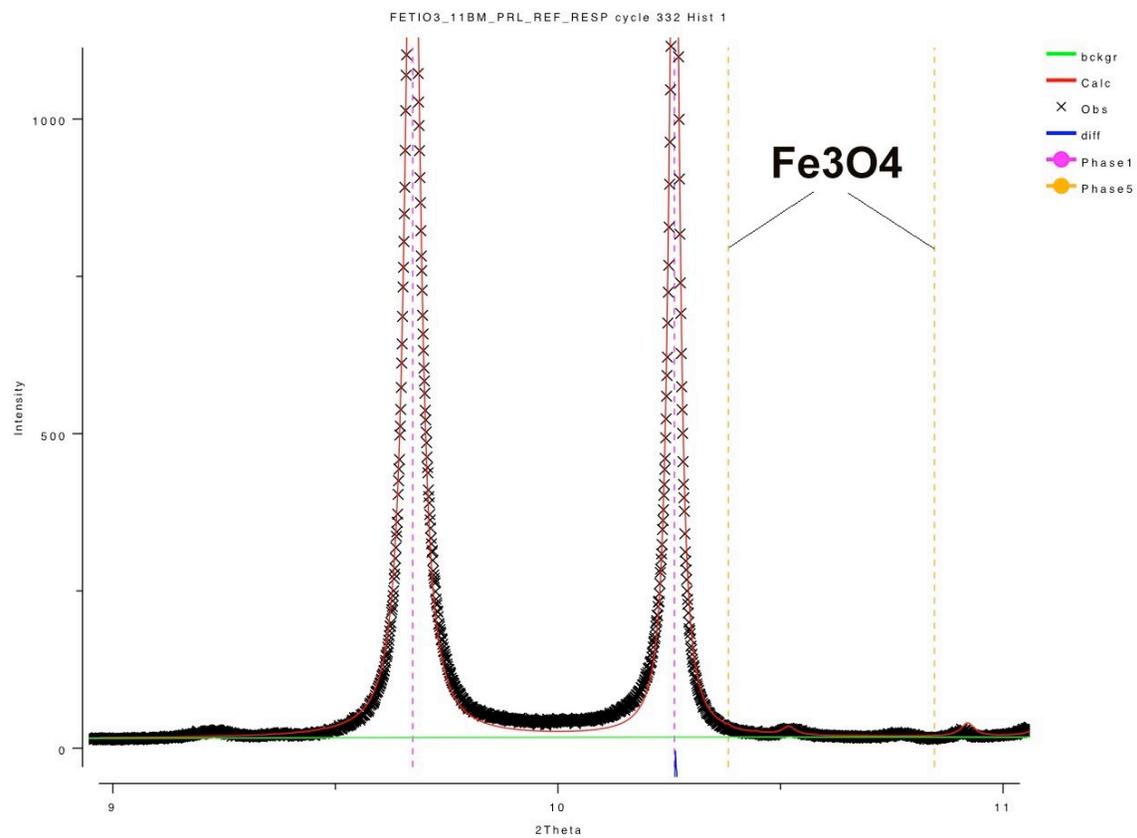

S18. Rietveld fit including Fe$_3$O$_4$ as a candidate impurity phase.

## S6. Piezoforce Microscopy: Frequency and Voltage Dependence

As discussed in the manuscript and detailed above in section S3, we prepared two samples of $FeTiO_3$-II. Fig. S19 shows a comparison of PFM data collected on the two samples. It is clear that the amplitude and phase maps of the two samples show the same features; the only marked difference is the grain size. As shown in section S3, this sample is also weakly ferromagnetic at ~120 K; thus, both samples are consistent and considering the analysis of impurity phases discussed above, we can reasonably conclude that they represent the intrinsic behavior of $FeTiO_3$-II.

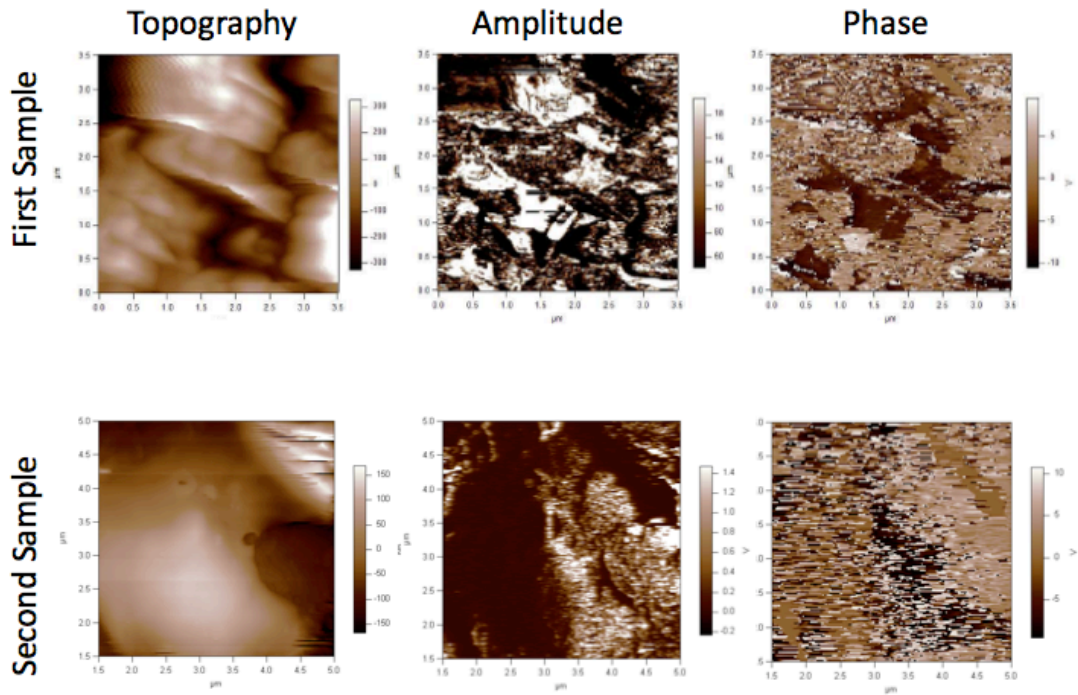

Fig. S19. Comparison of PFM data for two samples of $FeTiO_3$-II. Images are 3.5 x 3.5 $\mu m^2$.

Fig. S20 shows PFM data collected on ilmenite and on the second sample of FeTiO$_3$-II. The ac modulation voltage applied to the tip was 1.5 V$_{rms}$ at 17 kHz, and the applied force by the tip was 43 nN. The lack of contrast in ilmenite demonstrates it is not polar, while the contrast observed in amplitude and phase maps for the FeTiO$_3$-II sample show it is polar. It is important to note that we first imaged ilmenite and then using the same tip and voltage conditions, we imaged the FeTiO$_3$-II sample. This excludes the tip wear effect or other spurious artifacts involved in the measurement. If we reversed the order of measurement, one can argue that the tip has worn and it is not clear whether the lack of contrast in the ilmenite data is due to the non-polar nature or to a tip wear effect.

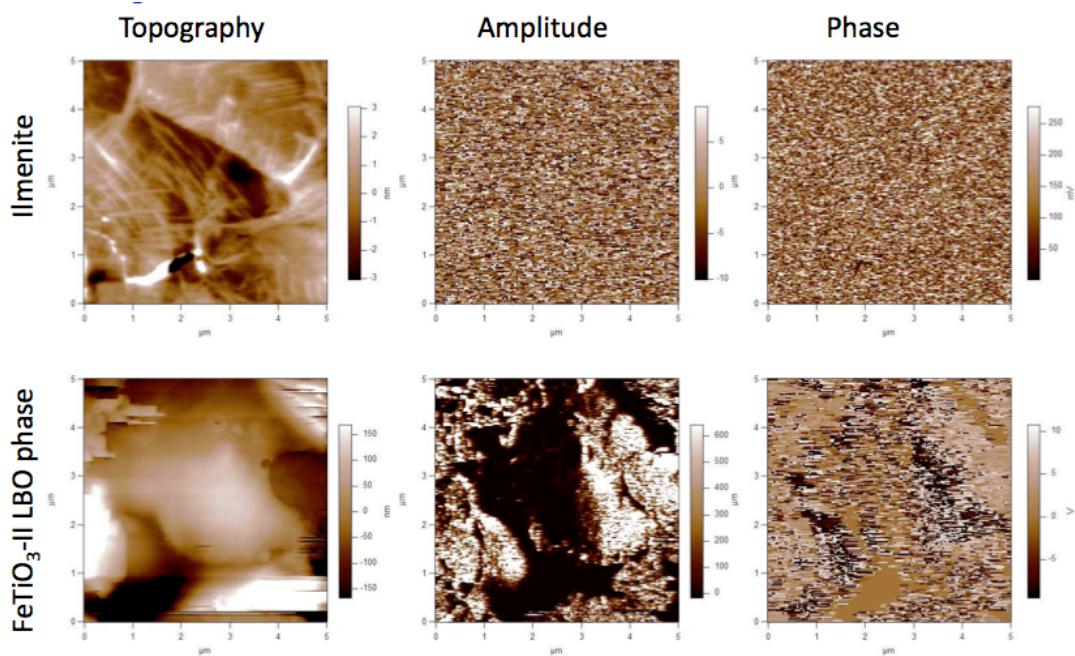

Fig. S20   Ilmenite and second FeTiO$_3$-II sample imaged using same tip and under identical voltage conditions (1.5 V$_{rms}$ at 17 KHz).  Images are 5 x 5 μm$^2$.

Fig S21 shows the effect of frequency on the PFM measurements on the second sample of $FeTiO_3$-II. Up to about 20 kHz, the amplitude and phase remain fairly constant; however, near the resonance peak at ~40 kHz, the amplitude shows a peak and phase reverses by 180°. We used 17 kHz as a reference frequency.

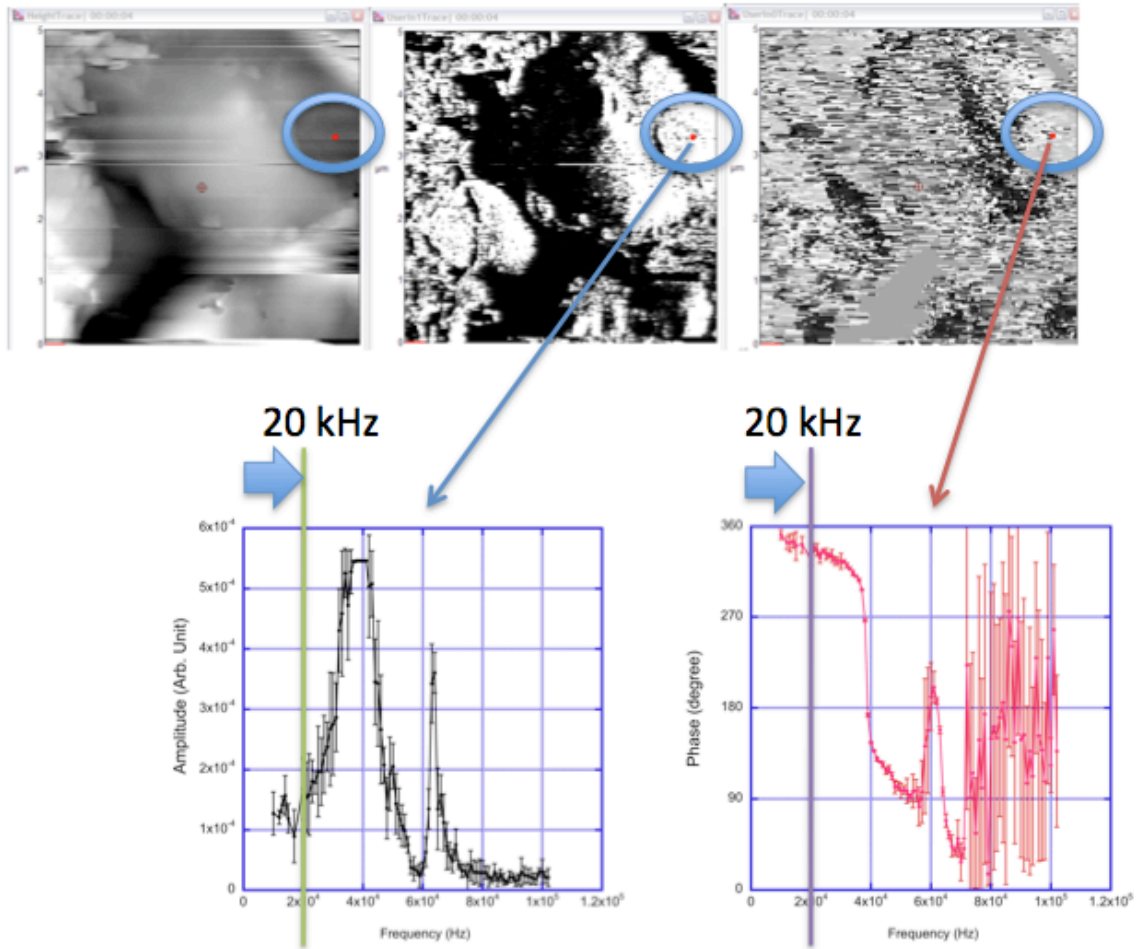

Fig. S21 Frequency dependence of PFM on the second $FeTiO_3$-II sample. Images are 5 x 5 $\mu m^2$.

Finally, in Fig. S22, we show the effect of changing the bias voltage of the tip on the PFM response. The data show no qualitative impact of changing the voltage from 0.5 $V_{rms}$ to 1.5 $V_{rms}$. Quantitatively, the amplitude scales with the voltage and the signal/noise of the phase signal improves. Both effects are consistent with an increase in the interaction volume at higher tip voltage.

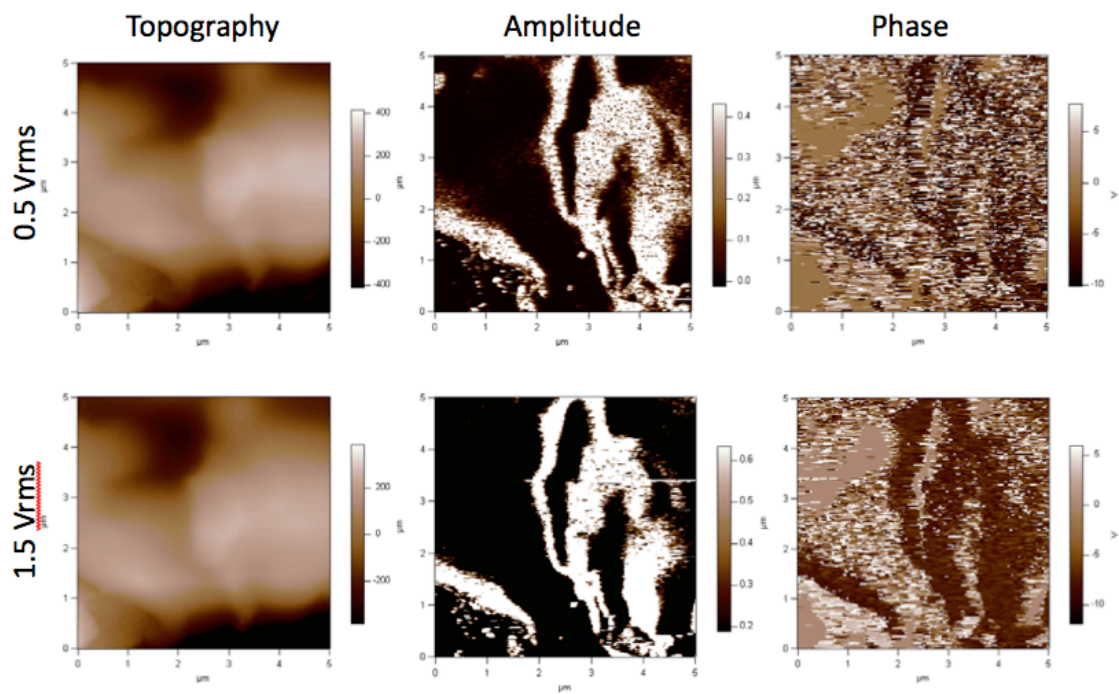

Fig. S22. Effect of voltage on PFM maps of the second FeTiO$_3$-II sample. Images are 5 x 5 μm$^2$.

# Appendix. Refined Crystal Structure Parameters

The results of the Rietveld refinement are tabulated in this appendix.

```
1FeTiO3 LiNbO3 phase TVFTHP2
PUBTABLE Version Win32  Mar 10 15:08:18 2009 Page   1

                |------------------------------------------|
                |      Program PUBTABLES Version Win32     |
                |   Generate crystal structure data tables |
                |   Distributed on Mon Nov 12 12:16:18 2007 |
                |------------------------------------------|

        |----------------------------------------------------------------
-|
        |              Allen C. Larson and Robert B. Von Dreele
|
        |         Manuel Lujan, Jr. Neutron Scattering Center, MS-H805
|
        |          Los Alamos National Laboratory, Los Alamos, NM  87545
|
        |
|
        | Copyright, 2000, The Regents of the University of California.
|
        |----------------------------------------------------------------
-|

 GENLES was run on Mar 10 14:59:58 2009    Total cycles run 323

 The Current Least-Squares controls are

 Maximum number of cycles is    3
 I/SigI cut-off is  1.00

 Anisotropic thermal factors are defined by
   T = exp(h**2*astr**2*u11+... +2*h*k*astr*bstr*u12+... )

 Space group R 3 c
 The lattice is acentric  R-centered trigonal     Laue symmetry 3barm1
  Multiplicity of a general site is  18
   The location of the origin is arbitrary in z

 The equivalent positions are:

 ( 1)     X       Y       Z  ( 2)    -Y     X-Y      Z  ( 3)    Y-X     -X
Z
 ( 4)    Y-X      Y    1/2+Z  ( 5)    -Y     -X    1/2+Z  ( 6)    X     X-Y
1/2+Z
```

```
 Lattice constants are
  a =  5.12362(4)  b = A  c =  13.74707(15)
  Alpha =  90     Beta =  90    Gamma =  120
 Cell volume = 312.532(5)

  Name         X              Y              Z         Ui/Ue*100    Site
sym Mult Type Seq Fractn
  Fe1       0.000000       0.000000       0.28802(9)      1.81         3
6   FE+2  1 1.012(6)
  Ti1       0.000000       0.000000       0.00076(9)      0.29         3
6   TI+4  2 0.980(5)
  O1        0.0523(4)      0.33982(34)    0.06242(16)     1.21(8)      1
18  O-2   3 1.0000

 Thermal parameters multiplied by 100.0 are
  Name       U11            U22            U33            U12            U13
U23
  Fe1       1.81           1.81           1.81           0.00           0.00
0.00
  Ti1       0.29           0.29           0.29           0.00           0.00
0.00
  O1        1.21(8)        1.21           1.21           0.00           0.00
0.00
1FeTiO3 LiNbO3 phase TVFTHP2
PUBTABLE Version Win32  Mar 10 15:08:18 2009 Page    2

 Space group F m -3 m
 The lattice is centric  F-centered cubic           Laue symmetry m3m
  Multiplicity of a general site is 192
  The symmetry of the point 0,0,0 contains 1bar

 The equivalent positions are:

 ( 1)    X       Y       Z    ( 2)    Z       X       Y    ( 3)    Y       Z
X
 ( 4)    X       Y      -Z    ( 5)   -Z       X       Y    ( 6)    Y      -Z
X
 ( 7)   -Z       X      -Y    ( 8)   -Y      -Z       X    ( 9)    Y      -Z
-X
 (10)   -X       Y      -Z    (11)   -Z      -X       Y    (12)    X      -Y
-Z
 (13)    Y       X       Z    (14)    Z       Y       X    (15)    X       Z
Y
 (16)    Y       X      -Z    (17)   -Z       Y       X    (18)    X      -Z
Y
 (19)   -Z       Y      -X    (20)   -X      -Z       Y    (21)    X      -Z
-Y
 (22)   -Y       X      -Z    (23)   -Z      -Y       X    (24)    Y      -X
-Z

 Lattice constants are
  a =  4.08579(7)  b = A  c = A
  Alpha =  90     Beta =  90    Gamma =  90
 Cell volume = 68.2071(19)
```

```
  Name           X            Y            Z         Ui/Ue*100    Site
sym Mult Type Seq Fractn
  Ti1        0.000000     0.000000     0.000000       0.60
M3M     4    TI    1 1.0000

 Thermal parameters multiplied by 100.0 are
  Name         U11          U22          U33          U12          U13
U23
  Ti1         0.60         0.60         0.60         0.00         0.00
0.00
```
1FeTiO3 LiNbO3 phase TVFTHP2
PUBTABLE Version Win32  Mar 10 15:08:18 2009 Page   3

```
 Space group I m -3 m
 The lattice is centric   I-centered cubic        Laue symmetry m3m
  Multiplicity of a general site is   96
  The symmetry of the point 0,0,0 contains 1bar

 The equivalent positions are:

 ( 1)     X        Y        Z  ( 2)     Z        X        Y  ( 3)     Y        Z
X
 ( 4)     X        Y       -Z  ( 5)    -Z        X        Y  ( 6)     Y       -Z
X
 ( 7)    -Z        X       -Y  ( 8)    -Y       -Z        X  ( 9)     Y       -Z
-X
 (10)    -X        Y       -Z  (11)    -Z       -X        Y  (12)     X       -Y
-Z
 (13)     Y        X        Z  (14)     Z        Y        X  (15)     X        Z
Y
 (16)     Y        X       -Z  (17)    -Z        Y        X  (18)     X       -Z
Y
 (19)    -Z        Y       -X  (20)    -X       -Z        Y  (21)     X       -Z
-Y
 (22)    -Y        X       -Z  (23)    -Z       -Y        X  (24)     Y       -X
-Z

 Lattice constants are
  a =  2.77454(8)  b =    A   c =    A
  Alpha =  90     Beta =  90    Gamma =  90
 Cell volume =  21.3587(11)

  Name           X            Y            Z         Ui/Ue*100    Site
sym Mult Type Seq Fractn
  Fe1        0.000000     0.000000     0.000000       0.22
M3M     2    FE    1 1.0000

 Thermal parameters multiplied by 100.0 are
  Name         U11          U22          U33          U12          U13
U23
  Fe1         0.22         0.22         0.22         0.00         0.00
0.00
```
1FeTiO3 LiNbO3 phase TVFTHP2
PUBTABLE Version Win32  Mar 10 15:08:18 2009 Page   4

```
 Space group F m -3 m
 The lattice is centric  F-centered cubic      Laue symmetry m3m
  Multiplicity of a general site is 192
  The symmetry of the point 0,0,0 contains 1bar

 The equivalent positions are:

 ( 1)     X        Y        Z   ( 2)    Z        X        Y   ( 3)    Y        Z
X
 ( 4)     X        Y       -Z   ( 5)   -Z        X        Y   ( 6)    Y       -Z
X
 ( 7)    -Z        X       -Y   ( 8)   -Y       -Z        X   ( 9)    Y       -Z
-X
 (10)    -X        Y       -Z   (11)   -Z       -X        Y   (12)    X       -Y
-Z
 (13)     Y        X        Z   (14)    Z        Y        X   (15)    X        Z
Y
 (16)     Y        X       -Z   (17)   -Z        Y        X   (18)    X       -Z
Y
 (19)    -Z        Y       -X   (20)   -X       -Z        Y   (21)    X       -Z
-Y
 (22)    -Y        X       -Z   (23)   -Z       -Y        X   (24)    Y       -X
-Z

 Lattice constants are
  a =  4.32902(30)  b = A   c = A
  Alpha =  90    Beta =  90    Gamma =  90
 Cell volume = 81.128(10)

  Name          X             Y             Z         Ui/Ue*100     Site
sym Mult Type Seq Fractn
  Fe1       0.000000      0.000000      0.000000        0.55
M3M     4   FE+2  1 1.0000
  O1        0.500000      0.500000      0.500000        0.55
M3M     4   O-2   2 1.0000

 Thermal parameters multiplied by 100.0 are
  Name       U11           U22           U33           U12           U13
U23
  Fe1       0.55          0.55          0.55          0.00          0.00
0.00
  O1        0.55          0.55          0.55          0.00          0.00
0.00
1FeTiO3 LiNbO3 phase TVFTHP2
PUBTABLE Version Win32  Mar 10 15:08:18 2009 Page    5

 Space group P -3 m 1
 The lattice is centric   primitive trigonal    Laue symmetry 3barm1
  Multiplicity of a general site is  12
  The symmetry of the point 0,0,0 contains 1bar

 The equivalent positions are:
```

```
 ( 1)    X       Y      Z ( 2)    -Y      X-Y     Z ( 3)   Y-X     -X
Z
 ( 4)    Y-X     Y      Z ( 5)    -Y      -X      Z ( 6)   X       X-Y
Z

 Lattice constants are
  a =  2.96448(16)  b = A   c =  4.81540(34)
  Alpha =  90     Beta =  90    Gamma =  120
 Cell volume =  36.649(4)

  Name          X             Y             Z          Ui/Ue*100    Site
sym Mult Type Seq Fractn
  Ti1      0.333330      0.666660      0.250000        2.50
3M(100)   2   TI    1 1.0000
  O1       0.000000      0.000000      0.000000        2.50          -
3M(100)   1   O     2 1.0000

 Thermal parameters multiplied by 100.0 are
  Name           U11           U22           U33           U12           U13
U23
  Ti1       2.50          2.50          2.50          0.00          0.00
0.00
  O1        2.50          2.50          2.50          0.00          0.00
0.00
1FeTiO3 LiNbO3 phase TVFTHP2
PUBTABLE Version Win32  Mar 10 15:08:18 2009 Page    6

 Space group C m c m
 The lattice is centric  C-centered orthorhombic  Laue symmetry mmm
  Multiplicity of a general site is   16
  The symmetry of the point 0,0,0 contains 1bar

  The equivalent positions are:

 ( 1)    X       Y      Z ( 2)    -X      Y       Z
 ( 3)    X      -Y    1/2+Z ( 4)    -X     -Y    1/2+Z

 Lattice constants are
  a =  3.7479(29)  b =  9.790(10)  c =  10.110(17)
  Alpha =  90     Beta =  90    Gamma =  90
 Cell volume =  371.0(8)

  Name          X             Y             Z          Ui/Ue*100    Site
sym Mult Type Seq Fractn
  Fe1      0.000000      0.136000      0.564200       46.00
M(100)    8   FE+3  1 0.6667
  Ti1      0.000000      0.136000      0.564200       46.00
M(100)    8   TI+4  2 0.3333
  Fe2      0.000000      0.189000      0.250000       37.00
MM2(010)  4   FE+3  3 0.6667
  Ti2      0.000000      0.189000      0.250000       37.00
MM2(010)  4   TI+4  4 0.3333
  O1       0.000000      0.766000      0.250000       73.00
MM2(010)  4   O-2   5 1.0000
```

```
    O2          0.000000         0.048000          0.117000        62.00
M(100)    8    O-2    6 1.0000
    O3          0.000000         0.311000          0.070000        66.00
M(100)    8    O-2    7 1.0000

 Thermal parameters multiplied by 100.0 are
   Name          U11              U22               U33             U12             U13
U23
   Fe1          46.00            46.00             46.00           0.00            0.00
0.00
   Ti1          46.00            46.00             46.00           0.00            0.00
0.00
   Fe2          37.00            37.00             37.00           0.00            0.00
0.00
   Ti2          37.00            37.00             37.00           0.00            0.00
0.00
   O1           73.00            73.00             73.00           0.00            0.00
0.00
   O2           62.00            62.00             62.00           0.00            0.00
0.00
   O3           66.00            66.00             66.00           0.00            0.00
0.00
```

**Bonding distances and angles from the Rietveld refinement for each phase present
(see Table S1)**

1FeTiO3 LiNbO3 phase TVFTHP2               Version    Mar 10 15:13:40 2009 Page   1

```
        |----------------------------------------------|
        |        Program DISAGL Version Win32          |
        | Crystal structure distance and angle program |
        |   Distributed on Mon Nov 12 12:15:40 2007    |
        |----------------------------------------------|

    |--------------------------------------------------------------|
    |        Allen C. Larson and Robert B. Von Dreele              |
    |     Manuel Lujan, Jr. Neutron Scattering Center, MS-H805     |
    |       Los Alamos National Laboratory, Los Alamos, NM  87545  |
    |                                                              |
    | Copyright, 2000, The Regents of the University of California.|
    |--------------------------------------------------------------|
```

The last history record is :
   HSTRY 43 GENLES  Win32  Mar 10 14:59:49 2009 Sdsq= 0.242E+06 S/E= 0.964E-02

Space group R 3 c
The lattice is acentric  R-centered trigonal     Laue symmetry 3barm1
Multiplicity of a general site is  18
The location of the origin is arbitrary in z

The equivalent positions are:

( 1)    X     Y     Z  ( 2)   -Y    X-Y   Z  ( 3)   Y-X   -X    Z
( 4)   Y-X    Y   1/2+Z ( 5)   -Y    -X   1/2+Z ( 6)   X    X-Y  1/2+Z

GENLES was run on Mar 10 14:59:58 2009    Total cycles run 323

The Current Least-Squares controls are

Maximum number of cycles is    3
I/SigI cut-off is  1.00

The atom type radii are

```
TYPE    BOND    ANGLE
FE+2   1.47000  1.27000
TI+4   1.66000  1.46000
O-2    1.09000  0.89000
TI     1.66000  1.46000
FE     1.47000  1.27000
O      1.09000  0.89000
FE+3   1.47000  1.27000
```

Lattice constants are   5.12362   5.12362  13.74707 0.00000 0.00000-0.50000
  Standard deviations   0.00004   0.00004   0.00015 0.00000 0.00000 0.00000

DMAX is   0.00000 DAGL is   0.00000 and IAGL is    0

The atoms read in are
1FeTiO3 LiNbO3 phase TVFTHP2                                 DISAGL   Version Win32
Mar 10 15:13:40 2009 Page   2

```
 Fe1    0.0000000(0)  0.0000000(0)  0.28802(9)    3    FE+2   1 1.012(6)
 Ti1    0.0000000(0)  0.0000000(0)  0.00076(9)    3    TI+4   2 0.980(5)
 O1     0.0523(4)     0.33982(34)   0.06242(16)   1    O-2    3 1.000(0)
```
1FeTiO3 LiNbO3 phase TVFTHP2                                 DISAGL   Version Win32
Mar 10 15:13:40 2009 Page   3

```
    Vector       Length       Optr Cell      Neighbor atom coordinates
    Fe1_Ti1      2.9246(8)      4 0 0 0     0.00000  0.00000  0.50076
    Fe1_Ti1      3.02519(18)  201-1-1 0    -0.33333 -0.66667  0.33410
    Fe1_Ti1      3.02519(18)  201-1 0 0    -0.33333  0.33333  0.33410
    Fe1_Ti1      3.02519(18)  201 0 0 0     0.66667  0.33333  0.33410
    Fe1_O1       2.1205(26)   104-1-1-1    -0.37919  0.00649  0.22909
    Fe1_O1       2.1205(26)   105 0-1-1    -0.00649 -0.38567  0.22909
    Fe1_O1       2.1205(26)   106 0 0-1     0.38567  0.37919  0.22909
    Fe1_O1       2.1587(26)   201-1-1 0    -0.28099 -0.32685  0.39576
    Fe1_O1       2.1587(26)   202 0 0 0     0.32685  0.04585  0.39576
    Fe1_O1       2.1587(26)   203-1 0 0    -0.04585  0.28099  0.39576

       Angle       Degrees     atom 1 loc  atom 3 loc
      O1_Fe1_O1    106.33(9)      104-1-1-1   105 0-1-1
      O1_Fe1_O1    106.33(9)      104-1-1-1   106 0 0-1
      O1_Fe1_O1     90.77(5)      104-1-1-1   201-1-1 0
      O1_Fe1_O1    158.03(14)     104-1-1-1   202 0 0 0
      O1_Fe1_O1     81.12(5)      104-1-1-1   203-1 0 0
      O1_Fe1_O1    106.33(9)      105 0-1-1   106 0 0-1
      O1_Fe1_O1     81.12(5)      105 0-1-1   201-1-1 0
      O1_Fe1_O1     90.77(5)      105 0-1-1   202 0 0 0
```

| | | | |
|---|---|---|---|
| O1_Fe1_O1 | 158.03(14) | 105 0 -1 -1 | 203 -1 0 0 |
| O1_Fe1_O1 | 158.03(14) | 106 0 0 -1 | 201 -1 -1 0 |
| O1_Fe1_O1 | 81.12(5) | 106 0 0 -1 | 202 0 0 0 |
| O1_Fe1_O1 | 90.77(5) | 106 0 0 -1 | 203 -1 0 0 |
| O1_Fe1_O1 | 78.10(11) | 201 -1 -1 0 | 202 0 0 0 |
| O1_Fe1_O1 | 78.10(11) | 201 -1 -1 0 | 203 -1 0 0 |
| O1_Fe1_O1 | 78.10(11) | 202 0 0 0 | 203 -1 0 0 |

| Vector | Length | Optr Cell | Neighbor atom coordinates |
|---|---|---|---|
| Ti1_Fe1 | 2.9246(8) | 4 0 0 -1 | 0.00000  0.00000 -0.21198 |
| Ti1_Fe1 | 3.02519(18) | 101 -1 -1 -1 | -0.66667 -0.33333 -0.04532 |
| Ti1_Fe1 | 3.02519(18) | 101 0 -1 -1 | 0.33333 -0.33333 -0.04532 |
| Ti1_Fe1 | 3.02519(18) | 101 0 0 -1 | 0.33333  0.66667 -0.04532 |
| Ti1_O1 | 1.8317(24) | 1 0 0 0 | 0.05234  0.33982  0.06242 |
| Ti1_O1 | 1.8317(24) | 2 0 0 0 | -0.33982 -0.28748  0.06242 |
| Ti1_O1 | 1.8317(24) | 3 0 0 0 | 0.28748 -0.05234  0.06242 |
| Ti1_O1 | 2.1330(25) | 204 -1 -1 -1 | -0.04585 -0.32685 -0.10424 |
| Ti1_O1 | 2.1330(25) | 205 0 0 -1 | 0.32685  0.28099 -0.10424 |
| Ti1_O1 | 2.1330(25) | 206 -1 0 -1 | -0.28099  0.04585 -0.10424 |

| Angle | Degrees | atom 1 loc | atom 3 loc |
|---|---|---|---|
| O1_Ti1_O1 | 100.30(12) | 1 0 0 0 | 2 0 0 0 |
| O1_Ti1_O1 | 100.30(12) | 1 0 0 0 | 3 0 0 0 |
| O1_Ti1_O1 | 164.97(17) | 1 0 0 0 | 204 -1 -1 -1 |
| O1_Ti1_O1 | 88.81(4) | 1 0 0 0 | 205 0 0 -1 |
| O1_Ti1_O1 | 89.69(4) | 1 0 0 0 | 206 -1 0 -1 |
| O1_Ti1_O1 | 100.30(12) | 2 0 0 0 | 3 0 0 0 |
| O1_Ti1_O1 | 89.69(4) | 2 0 0 0 | 204 -1 -1 -1 |
| O1_Ti1_O1 | 164.97(17) | 2 0 0 0 | 205 0 0 -1 |
| O1_Ti1_O1 | 88.81(4) | 2 0 0 0 | 206 -1 0 -1 |
| O1_Ti1_O1 | 88.81(4) | 3 0 0 0 | 204 -1 -1 -1 |
| O1_Ti1_O1 | 89.69(4) | 3 0 0 0 | 205 0 0 -1 |
| O1_Ti1_O1 | 164.97(17) | 3 0 0 0 | 206 -1 0 -1 |
| O1_Ti1_O1 | 79.23(11) | 204 -1 -1 -1 | 205 0 0 -1 |
| O1_Ti1_O1 | 79.23(11) | 204 -1 -1 -1 | 206 -1 0 -1 |
| O1_Ti1_O1 | 79.23(11) | 205 0 0 -1 | 206 -1 0 -1 |

| Vector | Length | Optr Cell | Neighbor atom coordinates |
|---|---|---|---|
| O1_Fe1 | 2.1587(26) | 101 0 0 -1 | 0.33333  0.66667 -0.04532 |
| O1_Fe1 | 2.1205(26) | 204 -1 0 -1 | -0.33333  0.33333  0.12135 |
| O1_Ti1 | 1.8317(24) | 1 0 0 0 | 0.00000  0.00000  0.00076 |
| O1_Ti1 | 2.1330(25) | 104 0 0 -1 | 0.33333  0.66667  0.16743 |

```
     Angle          Degrees       atom 1 loc  atom 3 loc
   Fe1_O1_Fe1       121.94(9)      101 0 0-1   204-1 0-1
   Fe1_O1_Ti1        98.26(15)     101 0 0-1    1 0 0 0
   Fe1_O1_Ti1        85.91(5)      101 0 0-1   104 0 0-1
   Fe1_O1_Ti1       118.00(7)      204-1 0-1    1 0 0 0
   Fe1_O1_Ti1        90.67(12)     204-1 0-1   104 0 0-1
   Ti1_O1_Ti1       141.26(11)      1 0 0 0    104 0 0-1
```
1FeTiO3 LiNbO3 phase TVFTHP2                        DISAGL   Version Win32
Mar 10 15:13:40 2009 Page   5

Space group F m -3 m
The lattice is centric  F-centered cubic       Laue symmetry m3m
 Multiplicity of a general site is 192
 The symmetry of the point 0,0,0 contains 1bar

The equivalent positions are:

```
( 1)   X    Y    Z  ( 2)   Z    X    Y  ( 3)   Y    Z    X
( 4)   X    Y   -Z  ( 5)  -Z    X    Y  ( 6)   Y   -Z    X
( 7)  -Z    X   -Y  ( 8)  -Y   -Z    X  ( 9)   Y   -Z   -X
(10)  -X    Y   -Z  (11)  -Z   -X    Y  (12)   X   -Y   -Z
(13)   Y    X    Z  (14)   Z    Y    X  (15)   X    Z    Y
(16)   Y    X   -Z  (17)  -Z    Y    X  (18)   X   -Z    Y
(19)  -Z    Y   -X  (20)  -X   -Z    Y  (21)   X   -Z   -Y
(22)  -Y    X   -Z  (23)  -Z   -Y    X  (24)   Y   -X   -Z
```

GENLES was run on Mar 10 14:59:58 2009    Total cycles run 323

The Current Least-Squares controls are

Maximum number of cycles is    3
I/SigI cut-off is  1.00

The atom type radii are

```
TYPE    BOND     ANGLE
FE+2    1.47000  1.27000
TI+4    1.66000  1.46000
O-2     1.09000  0.89000
TI      1.66000  1.46000
FE      1.47000  1.27000
O       1.09000  0.89000
FE+3    1.47000  1.27000
```

Lattice constants are    4.08579   4.08579   4.08579 0.00000 0.00000 0.00000

Standard deviations   0.00007  0.00007  0.00007 0.00000 0.00000 0.00000

DMAX is   2.18000 DAGL is   1.78000 and IAGL is   0

The atoms read in are
 Ti1    0.0000000(0) 0.0000000(0) 0.0000000(0)   M3M    TI    1 1.000(0)
1FeTiO3 LiNbO3 phase TVFTHP2                              DISAGL   Version Win32
Mar 10 15:13:40 2009 Page   6

|   Vector   |   Length    |  Optr Cell   | Neighbor atom coordinates |
|------------|-------------|--------------|---------------------------|
| Ti1_Ti1    | 2.88909(3)  | 101 0-1-1    |  0.00000 -0.50000 -0.50000 |
| Ti1_Ti1    | 2.88909(3)  | 101 0-1 0    |  0.00000 -0.50000  0.50000 |
| Ti1_Ti1    | 2.88909(3)  | 101 0 0-1    |  0.00000  0.50000 -0.50000 |
| Ti1_Ti1    | 2.88909(3)  | 101 0 0 0    |  0.00000  0.50000  0.50000 |
| Ti1_Ti1    | 2.88909(3)  | 201-1 0-1    | -0.50000  0.00000 -0.50000 |
| Ti1_Ti1    | 2.88909(3)  | 201-1 0 0    | -0.50000  0.00000  0.50000 |
| Ti1_Ti1    | 2.88909(3)  | 201 0 0-1    |  0.50000  0.00000 -0.50000 |
| Ti1_Ti1    | 2.88909(3)  | 201 0 0 0    |  0.50000  0.00000  0.50000 |
| Ti1_Ti1    | 2.88909(5)  | 301-1-1 0    | -0.50000 -0.50000  0.00000 |
| Ti1_Ti1    | 2.88909(5)  | 301-1 0 0    | -0.50000  0.50000  0.00000 |
| Ti1_Ti1    | 2.88909(5)  | 301 0-1 0    |  0.50000 -0.50000  0.00000 |
| Ti1_Ti1    | 2.88909(5)  | 301 0 0 0    |  0.50000  0.50000  0.00000 |

|    Angle     |  Degrees   | atom 1 loc | atom 3 loc |
|--------------|------------|------------|------------|
| Ti1_Ti1_Ti1  |  90.000(1) | 101 0-1-1  | 101 0-1 0  |
| Ti1_Ti1_Ti1  |  90.000(1) | 101 0-1-1  | 101 0 0-1  |
| Ti1_Ti1_Ti1  | 179.972(0) | 101 0-1-1  | 101 0 0 0  |
| Ti1_Ti1_Ti1  |  60.000(1) | 101 0-1-1  | 201-1 0-1  |
| Ti1_Ti1_Ti1  | 120.000(1) | 101 0-1-1  | 201-1 0 0  |
| Ti1_Ti1_Ti1  |  60.000(1) | 101 0-1-1  | 201 0 0-1  |
| Ti1_Ti1_Ti1  | 120.000(1) | 101 0-1-1  | 201 0 0 0  |
| Ti1_Ti1_Ti1  |  60.000(0) | 101 0-1-1  | 301-1-1 0  |
| Ti1_Ti1_Ti1  | 120.000(0) | 101 0-1-1  | 301-1 0 0  |
| Ti1_Ti1_Ti1  |  60.000(0) | 101 0-1-1  | 301 0-1 0  |
| Ti1_Ti1_Ti1  | 120.000(0) | 101 0-1-1  | 301 0 0 0  |
| Ti1_Ti1_Ti1  | 179.966(0) | 101 0-1 0  | 101 0 0-1  |
| Ti1_Ti1_Ti1  |  90.000(1) | 101 0-1 0  | 101 0 0 0  |
| Ti1_Ti1_Ti1  | 120.000(1) | 101 0-1 0  | 201-1 0-1  |
| Ti1_Ti1_Ti1  |  60.000(1) | 101 0-1 0  | 201-1 0 0  |
| Ti1_Ti1_Ti1  | 120.000(1) | 101 0-1 0  | 201 0 0-1  |
| Ti1_Ti1_Ti1  |  60.000(1) | 101 0-1 0  | 201 0 0 0  |
| Ti1_Ti1_Ti1  |  60.000(0) | 101 0-1 0  | 301-1-1 0  |
| Ti1_Ti1_Ti1  | 120.000(0) | 101 0-1 0  | 301-1 0 0  |
| Ti1_Ti1_Ti1  |  60.000(0) | 101 0-1 0  | 301 0-1 0  |
| Ti1_Ti1_Ti1  | 120.000(0) | 101 0-1 0  | 301 0 0 0  |

| Angle | Degrees | atom 1 loc | atom 3 loc |
|---|---|---|---|
| Ti1_Ti1_Ti1 | 90.000(1) | 101 0 0 -1 | 101 0 0 0 |
| Ti1_Ti1_Ti1 | 60.000(1) | 101 0 0 -1 | 201 -1 0 -1 |
| Ti1_Ti1_Ti1 | 120.000(1) | 101 0 0 -1 | 201 -1 0 0 |
| Ti1_Ti1_Ti1 | 60.000(1) | 101 0 0 -1 | 201 0 0 -1 |
| Ti1_Ti1_Ti1 | 120.000(1) | 101 0 0 -1 | 201 0 0 0 |
| Ti1_Ti1_Ti1 | 120.000(0) | 101 0 0 -1 | 301 -1 -1 0 |
| Ti1_Ti1_Ti1 | 60.000(0) | 101 0 0 -1 | 301 -1 0 0 |
| Ti1_Ti1_Ti1 | 120.000(0) | 101 0 0 -1 | 301 0 -1 0 |
| Ti1_Ti1_Ti1 | 60.000(0) | 101 0 0 -1 | 301 0 0 0 |
| Ti1_Ti1_Ti1 | 120.000(1) | 101 0 0 0 | 201 -1 0 -1 |
| Ti1_Ti1_Ti1 | 60.000(1) | 101 0 0 0 | 201 -1 0 0 |
| Ti1_Ti1_Ti1 | 120.000(1) | 101 0 0 0 | 201 0 0 -1 |
| Ti1_Ti1_Ti1 | 60.000(1) | 101 0 0 0 | 201 0 0 0 |
| Ti1_Ti1_Ti1 | 120.000(0) | 101 0 0 0 | 301 -1 -1 0 |
| Ti1_Ti1_Ti1 | 60.000(0) | 101 0 0 0 | 301 -1 0 0 |
| Ti1_Ti1_Ti1 | 120.000(0) | 101 0 0 0 | 301 0 -1 0 |
| Ti1_Ti1_Ti1 | 60.000(0) | 101 0 0 0 | 301 0 0 0 |
| Ti1_Ti1_Ti1 | 90.000(1) | 201 -1 0 -1 | 201 -1 0 0 |
| Ti1_Ti1_Ti1 | 90.000(1) | 201 -1 0 -1 | 201 0 0 -1 |
| Ti1_Ti1_Ti1 | 179.972(0) | 201 -1 0 -1 | 201 0 0 0 |
| Ti1_Ti1_Ti1 | 60.000(0) | 201 -1 0 -1 | 301 -1 -1 0 |

1FeTiO3 LiNbO3 phase TVFTHP2         DISAGL   Version Win32
Mar 10 15:13:40 2009 Page   7

| Angle | Degrees | atom 1 loc | atom 3 loc |
|---|---|---|---|
| Ti1_Ti1_Ti1 | 60.000(0) | 201 -1 0 -1 | 301 -1 0 0 |
| Ti1_Ti1_Ti1 | 120.000(0) | 201 -1 0 -1 | 301 0 -1 0 |
| Ti1_Ti1_Ti1 | 120.000(0) | 201 -1 0 -1 | 301 0 0 0 |
| Ti1_Ti1_Ti1 | 179.966(0) | 201 -1 0 0 | 201 0 0 -1 |
| Ti1_Ti1_Ti1 | 90.000(1) | 201 -1 0 0 | 201 0 0 0 |
| Ti1_Ti1_Ti1 | 60.000(0) | 201 -1 0 0 | 301 -1 -1 0 |
| Ti1_Ti1_Ti1 | 60.000(0) | 201 -1 0 0 | 301 -1 0 0 |
| Ti1_Ti1_Ti1 | 120.000(0) | 201 -1 0 0 | 301 0 -1 0 |
| Ti1_Ti1_Ti1 | 120.000(0) | 201 -1 0 0 | 301 0 0 0 |
| Ti1_Ti1_Ti1 | 90.000(1) | 201 0 0 -1 | 201 0 0 0 |
| Ti1_Ti1_Ti1 | 120.000(0) | 201 0 0 -1 | 301 -1 -1 0 |
| Ti1_Ti1_Ti1 | 120.000(0) | 201 0 0 -1 | 301 -1 0 0 |
| Ti1_Ti1_Ti1 | 60.000(0) | 201 0 0 -1 | 301 0 -1 0 |
| Ti1_Ti1_Ti1 | 60.000(0) | 201 0 0 -1 | 301 0 0 0 |
| Ti1_Ti1_Ti1 | 120.000(0) | 201 0 0 0 | 301 -1 -1 0 |
| Ti1_Ti1_Ti1 | 120.000(0) | 201 0 0 0 | 301 -1 0 0 |
| Ti1_Ti1_Ti1 | 60.000(0) | 201 0 0 0 | 301 0 -1 0 |
| Ti1_Ti1_Ti1 | 60.000(0) | 201 0 0 0 | 301 0 0 0 |
| Ti1_Ti1_Ti1 | 90.000(0) | 301 -1 -1 0 | 301 -1 0 0 |
| Ti1_Ti1_Ti1 | 90.000(0) | 301 -1 -1 0 | 301 0 -1 0 |

```
    Ti1_Ti1_Ti1      179.972(0)    301 -1 -1  0   301  0  0  0
    Ti1_Ti1_Ti1      179.966(0)    301 -1  0  0   301  0 -1  0
    Ti1_Ti1_Ti1       90.000(0)    301 -1  0  0   301  0  0  0
    Ti1_Ti1_Ti1       90.000(0)    301  0 -1  0   301  0  0  0
```
1FeTiO3 LiNbO3 phase TVFTHP2                                    DISAGL   Version Win32
Mar 10 15:13:40 2009 Page   8

Space group I m -3 m
The lattice is centric  I-centered cubic      Laue symmetry m3m
 Multiplicity of a general site is  96
 The symmetry of the point 0,0,0 contains 1bar

The equivalent positions are:

```
( 1)   X    Y    Z  ( 2)   Z    X    Y  ( 3)   Y    Z    X
( 4)   X    Y   -Z  ( 5)  -Z    X    Y  ( 6)   Y   -Z    X
( 7)  -Z    X   -Y  ( 8)  -Y   -Z    X  ( 9)   Y   -Z   -X
(10)  -X    Y   -Z  (11)  -Z   -X    Y  (12)   X   -Y   -Z
(13)   Y    X    Z  (14)   Z    Y    X  (15)   X    Z    Y
(16)   Y    X   -Z  (17)  -Z    Y    X  (18)   X   -Z    Y
(19)  -Z    Y   -X  (20)  -X   -Z    Y  (21)   X   -Z   -Y
(22)  -Y    X   -Z  (23)  -Z   -Y    X  (24)   Y   -X   -Z
```

GENLES was run on Mar 10 14:59:58 2009    Total cycles run 323

The Current Least-Squares controls are

Maximum number of cycles is   3
I/SigI cut-off is  1.00

The atom type radii are

```
TYPE    BOND     ANGLE
FE+2   1.47000   1.27000
TI+4   1.66000   1.46000
O-2    1.09000   0.89000
TI     1.66000   1.46000
FE     1.47000   1.27000
O      1.09000   0.89000
FE+3   1.47000   1.27000
```

Lattice constants are   2.77454  2.77454  2.77454 0.00000 0.00000 0.00000
  Standard deviations   0.00008  0.00008  0.00008 0.00000 0.00000 0.00000

DMAX is   3.32000 DAGL is   2.92000 and IAGL is   0

The atoms read in are
  Fe1    0.0000000(0) 0.0000000(0) 0.0000000(0)   M3M   FE    1 1.000(0)
1FeTiO3 LiNbO3 phase TVFTHP2                            DISAGL   Version Win32
Mar 10 15:13:40 2009 Page   9

    Vector      Length      Optr Cell       Neighbor atom coordinates
   Fe1_Fe1     2.77454(8)    1-1 0 0    -1.00000  0.00000  0.00000
   Fe1_Fe1     2.77454(8)    1 0-1 0     0.00000 -1.00000  0.00000
   Fe1_Fe1     2.77454(8)    1 0 0-1     0.00000  0.00000 -1.00000
   Fe1_Fe1     2.77454(8)    1 0 0 1     0.00000  0.00000  1.00000
   Fe1_Fe1     2.77454(8)    1 0 1 0     0.00000  1.00000  0.00000
   Fe1_Fe1     2.77454(8)    1 1 0 0     1.00000  0.00000  0.00000
   Fe1_Fe1     2.40283(5)   101-1-1-1   -0.50000 -0.50000 -0.50000
   Fe1_Fe1     2.40283(5)   101-1-1 0   -0.50000 -0.50000  0.50000
   Fe1_Fe1     2.40283(5)   101-1 0-1   -0.50000  0.50000 -0.50000
   Fe1_Fe1     2.40283(5)   101-1 0 0   -0.50000  0.50000  0.50000
   Fe1_Fe1     2.40283(5)   101 0-1-1    0.50000 -0.50000 -0.50000
   Fe1_Fe1     2.40283(5)   101 0-1 0    0.50000 -0.50000  0.50000
   Fe1_Fe1     2.40283(5)   101 0 0-1    0.50000  0.50000 -0.50000
   Fe1_Fe1     2.40283(5)   101 0 0 0    0.50000  0.50000  0.50000

    Angle       Degrees     atom 1 loc  atom 3 loc
  Fe1_Fe1_Fe1    70.529(2)    101-1-1-1  101-1-1 0
  Fe1_Fe1_Fe1    70.529(1)    101-1-1-1  101-1 0-1
  Fe1_Fe1_Fe1   109.471(1)    101-1-1-1  101-1 0 0
  Fe1_Fe1_Fe1    70.529(1)    101-1-1-1  101 0-1-1
  Fe1_Fe1_Fe1   109.471(1)    101-1-1-1  101 0-1 0
  Fe1_Fe1_Fe1   109.471(2)    101-1-1-1  101 0 0-1
  Fe1_Fe1_Fe1   180.000(0)    101-1-1-1  101 0 0 0
  Fe1_Fe1_Fe1   109.471(1)    101-1-1 0  101-1 0-1
  Fe1_Fe1_Fe1    70.529(1)    101-1-1 0  101-1 0 0
  Fe1_Fe1_Fe1   109.471(1)    101-1-1 0  101 0-1-1
  Fe1_Fe1_Fe1    70.529(1)    101-1-1 0  101 0-1 0
  Fe1_Fe1_Fe1   180.000(0)    101-1-1 0  101 0 0-1
  Fe1_Fe1_Fe1   109.471(2)    101-1-1 0  101 0 0 0
  Fe1_Fe1_Fe1    70.529(2)    101-1 0-1  101-1 0 0
  Fe1_Fe1_Fe1   109.471(2)    101-1 0-1  101 0-1-1
  Fe1_Fe1_Fe1   180.000(0)    101-1 0-1  101 0-1 0
  Fe1_Fe1_Fe1    70.529(1)    101-1 0-1  101 0 0-1
  Fe1_Fe1_Fe1   109.471(1)    101-1 0-1  101 0 0 0
  Fe1_Fe1_Fe1   180.000(0)    101-1 0 0  101 0-1-1
  Fe1_Fe1_Fe1   109.471(2)    101-1 0 0  101 0-1 0
  Fe1_Fe1_Fe1   109.471(1)    101-1 0 0  101 0 0-1
  Fe1_Fe1_Fe1    70.529(1)    101-1 0 0  101 0 0 0

```
        Fe1_Fe1_Fe1      70.529(2)     101 0-1-1  101 0-1 0
        Fe1_Fe1_Fe1      70.529(1)     101 0-1-1  101 0 0-1
        Fe1_Fe1_Fe1     109.471(1)     101 0-1-1  101 0 0 0
        Fe1_Fe1_Fe1     109.471(1)     101 0-1 0  101 0 0-1
        Fe1_Fe1_Fe1      70.529(1)     101 0-1 0  101 0 0 0
        Fe1_Fe1_Fe1      70.529(2)     101 0 0-1  101 0 0 0
```
1FeTiO3 LiNbO3 phase TVFTHP2                           DISAGL   Version Win32
Mar 10 15:13:40 2009 Page  10

 Space group F m -3 m
 The lattice is centric  F-centered cubic      Laue symmetry m3m
 Multiplicity of a general site is 192
 The symmetry of the point 0,0,0 contains 1bar

The equivalent positions are:

```
( 1)   X    Y     Z ( 2)   Z    X    Y ( 3)   Y    Z    X
( 4)   X    Y    -Z ( 5)  -Z    X    Y ( 6)   Y   -Z    X
( 7)  -Z    X    -Y ( 8)  -Y   -Z    X ( 9)   Y   -Z   -X
(10)  -X    Y    -Z (11)  -Z   -X    Y (12)   X   -Y   -Z
(13)   Y    X     Z (14)   Z    Y    X (15)   X    Z    Y
(16)   Y    X    -Z (17)  -Z    Y    X (18)   X   -Z    Y
(19)  -Z    Y    -X (20)  -X   -Z    Y (21)   X   -Z   -Y
(22)  -Y    X    -Z (23)  -Z   -Y    X (24)   Y   -X   -Z
```

 GENLES was run on Mar 10 14:59:58 2009    Total cycles run 323

The Current Least-Squares controls are

Maximum number of cycles is    3
I/SigI cut-off is  1.00

The atom type radii are

```
TYPE    BOND     ANGLE
FE+2   1.47000   1.27000
TI+4   1.66000   1.46000
O-2    1.09000   0.89000
TI     1.66000   1.46000
FE     1.47000   1.27000
O      1.09000   0.89000
FE+3   1.47000   1.27000
```

Lattice constants are    4.32902   4.32902   4.32902 0.00000 0.00000 0.00000
  Standard deviations    0.00030   0.00030   0.00030 0.00000 0.00000 0.00000

DMAX is 2.94000 DAGL is 2.54000 and IAGL is 0

The atoms read in are
Fe1   0.0000000(0) 0.0000000(0) 0.0000000(0)   M3M   FE+2   1 1.000(0)
O1    0.5000000(0) 0.5000000(0) 0.5000000(0)   M3M   O-2    2 1.000(0)
1FeTiO3 LiNbO3 phase TVFTHP2                              DISAGL   Version Win32
Mar 10 15:13:40 2009 Page  11

| Vector  | Length      | Optr Cell  | Neighbor atom coordinates |
|---------|-------------|------------|---------------------------|
| Fe1_O1  | 2.16451(15) | 101-1-1-1  | -0.50000  0.00000  0.00000 |
| Fe1_O1  | 2.16451(15) | 101 0-1-1  |  0.50000  0.00000  0.00000 |
| Fe1_O1  | 2.16451(15) | 201-1-1-1  |  0.00000 -0.50000  0.00000 |
| Fe1_O1  | 2.16451(15) | 201-1 0-1  |  0.00000  0.50000  0.00000 |
| Fe1_O1  | 2.16451(15) | 301-1-1-1  |  0.00000  0.00000 -0.50000 |
| Fe1_O1  | 2.16451(15) | 301-1-1 0  |  0.00000  0.00000  0.50000 |
| O1_Fe1  | 2.16451(15) | 101 0 0 0  |  0.00000  0.50000  0.50000 |
| O1_Fe1  | 2.16451(15) | 101 1 0 0  |  1.00000  0.50000  0.50000 |
| O1_Fe1  | 2.16451(15) | 201 0 0 0  |  0.50000  0.00000  0.50000 |
| O1_Fe1  | 2.16451(15) | 201 0 1 0  |  0.50000  1.00000  0.50000 |
| O1_Fe1  | 2.16451(15) | 301 0 0 0  |  0.50000  0.50000  0.00000 |
| O1_Fe1  | 2.16451(15) | 301 0 0 1  |  0.50000  0.50000  1.00000 |

1FeTiO3 LiNbO3 phase TVFTHP2                              DISAGL   Version Win32
Mar 10 15:13:40 2009 Page  12

Space group P -3 m 1
The lattice is centric   primitive trigonal     Laue symmetry 3barm1
Multiplicity of a general site is  12
The symmetry of the point 0,0,0 contains 1bar

The equivalent positions are:

( 1)   X     Y     Z ( 2)  -Y    X-Y   Z ( 3)  Y-X   -X    Z
( 4)  Y-X    Y     Z ( 5)  -Y    -X    Z ( 6)   X    X-Y   Z

GENLES was run on Mar 10 14:59:58 2009    Total cycles run 323

The Current Least-Squares controls are

Maximum number of cycles is   3
I/SigI cut-off is  1.00

The atom type radii are

```
TYPE    BOND    ANGLE
FE+2   1.47000  1.27000
TI+4   1.66000  1.46000
O-2    1.09000  0.89000
TI     1.66000  1.46000
FE     1.47000  1.27000
O      1.09000  0.89000
FE+3   1.47000  1.27000
```

Lattice constants are   2.96448  2.96448  4.81540 0.00000 0.00000-0.50000
 Standard deviations    0.00016  0.00016  0.00034 0.00000 0.00000 0.00000

DMAX is    2.18000 DAGL is    1.78000 and IAGL is    0

The atoms read in are
 Ti1    0.3333300(0)  0.6666600(0)  0.2500000(0)  3M(100)  TI     1 1.000(0)
 O1     0.0000000(0)  0.0000000(0)  0.0000000(0) -3M(100)  O      2 1.000(0)
1FeTiO3 LiNbO3 phase TVFTHP2                            DISAGL   Version Win32
Mar 10 15:13:40 2009 Page  13

       Vector       Length       Optr Cell      Neighbor atom coordinates
       Ti1_Ti1      2.96448(16)   1-1-1 0    -0.66667 -0.33334  0.25000
       Ti1_Ti1      2.96448(16)   1-1 0 0    -0.66667  0.66666  0.25000
       Ti1_Ti1      2.96448(16)   1 0-1 0     0.33333 -0.33334  0.25000
       Ti1_Ti1      2.96448(16)   1 0 1 0     0.33333  1.66666  0.25000
       Ti1_Ti1      2.96448(16)   1 1 0 0     1.33333  0.66666  0.25000
       Ti1_Ti1      2.96448(16)   1 1 1 0     1.33333  1.66666  0.25000
       Ti1_Ti1      2.95404(15)  -1 0 1 0    -0.33333  0.33334 -0.25000
       Ti1_Ti1      2.95404(15)  -1 0 1 1    -0.33333  0.33334  0.75000
       Ti1_Ti1      2.95404(15)  -1 1 1 0     0.66667  0.33334 -0.25000
       Ti1_Ti1      2.95404(15)  -1 1 1 1     0.66667  0.33334  0.75000
       Ti1_Ti1      2.95407(15)  -1 1 2 0     0.66667  1.33334 -0.25000
       Ti1_Ti1      2.95407(15)  -1 1 2 1     0.66667  1.33334  0.75000
       Ti1_O1       2.09250(9)    1 0 0 0     0.00000  0.00000  0.00000
       Ti1_O1       2.09253(9)    1 0 1 0     0.00000  1.00000  0.00000
       Ti1_O1       2.09253(9)    1 1 1 0     1.00000  1.00000  0.00000

       Angle      Degrees     atom 1 loc  atom 3 loc
       O1_Ti1_O1      90.202(3)     1 0 0 0    1 0 1 0
       O1_Ti1_O1      90.202(3)     1 0 0 0    1 1 1 0
       O1_Ti1_O1      90.202(3)     1 0 1 0    1 1 1 0

       Vector       Length       Optr Cell      Neighbor atom coordinates
       O1_Ti1       2.09253(9)    1-1-1 0    -0.66667 -0.33334  0.25000
       O1_Ti1       2.09253(9)    1 0-1 0     0.33333 -0.33334  0.25000

```
 O1_Ti1      2.09250(9)       1 0 0 0    0.33333  0.66666  0.25000
 O1_Ti1      2.09250(9)      -1 0 0 0   -0.33333 -0.66666 -0.25000
 O1_Ti1      2.09253(9)      -1 0 1 0   -0.33333  0.33334 -0.25000
 O1_Ti1      2.09253(9)      -1 1 1 0    0.66667  0.33334 -0.25000

    Angle         Degrees       atom 1 loc  atom 3 loc
  Ti1_O1_Ti1     90.202(3)      1 -1 -1 0    1 0 -1 0
  Ti1_O1_Ti1     90.202(3)      1 -1 -1 0    1 0 0 0
  Ti1_O1_Ti1     89.798(3)      1 -1 -1 0   -1 0 0 0
  Ti1_O1_Ti1     89.798(3)      1 -1 -1 0   -1 0 1 0
  Ti1_O1_Ti1    180.000(0)      1 -1 -1 0   -1 1 1 0
  Ti1_O1_Ti1     90.202(3)      1 0 -1 0    1 0 0 0
  Ti1_O1_Ti1     89.798(3)      1 0 -1 0   -1 0 0 0
  Ti1_O1_Ti1    180.000(0)      1 0 -1 0   -1 0 1 0
  Ti1_O1_Ti1     89.798(3)      1 0 -1 0   -1 1 1 0
  Ti1_O1_Ti1    180.000(0)      1 0 0 0    -1 0 0 0
  Ti1_O1_Ti1     89.798(3)      1 0 0 0    -1 0 1 0
  Ti1_O1_Ti1     89.798(3)      1 0 0 0    -1 1 1 0
  Ti1_O1_Ti1     90.202(3)     -1 0 0 0    -1 0 1 0
  Ti1_O1_Ti1     90.202(3)     -1 0 0 0    -1 1 1 0
  Ti1_O1_Ti1     90.202(3)     -1 0 1 0    -1 1 1 0
```

1FeTiO3 LiNbO3 phase TVFTHP2                    DISAGL   Version Win32
Mar 10 15:13:40 2009 Page  14

 Space group C m c m
 The lattice is centric  C-centered orthorhombic  Laue symmetry mmm
 Multiplicity of a general site is  16
 The symmetry of the point 0,0,0 contains 1bar

The equivalent positions are:

( 1)   X    Y    Z ( 2)  -X    Y    Z
( 3)   X   -Y  1/2+Z ( 4)  -X   -Y  1/2+Z

 GENLES was run on Mar 10 14:59:58 2009   Total cycles run 323

The Current Least-Squares controls are

Maximum number of cycles is   3
I/SigI cut-off is  1.00

The atom type radii are

TYPE    BOND    ANGLE
FE+2    1.47000  1.27000

```
TI+4    1.66000   1.46000
O-2     1.09000   0.89000
TI      1.66000   1.46000
FE      1.47000   1.27000
O       1.09000   0.89000
FE+3    1.47000   1.27000
```

Lattice constants are    3.74794   9.78973  10.11023  0.00000  0.00000  0.00000
 Standard deviations     0.00286   0.00993   0.01688  0.00000  0.00000  0.00000

DMAX is   2.18000 DAGL is    1.78000 and IAGL is    0

The atoms read in are
 Fe1    0.0000000(0)  0.1360000(0)  0.5642000(0)  M(100)   FE+3   1 0.667(0)
 Ti1    0.0000000(0)  0.1360000(0)  0.5642000(0)  M(100)   TI+4   2 0.333(0)
 Fe2    0.0000000(0)  0.1890000(0)  0.2500000(0)  MM2(010) FE+3   3 0.667(0)
 Ti2    0.0000000(0)  0.1890000(0)  0.2500000(0)  MM2(010) TI+4   4 0.333(0)
 O1     0.0000000(0)  0.7660000(0)  0.2500000(0)  MM2(010) O-2    5 1.000(0)
 O2     0.0000000(0)  0.0480000(0)  0.1170000(0)  M(100)   O-2    6 1.000(0)
 O3     0.0000000(0)  0.3110000(0)  0.0700000(0)  M(100)   O-2    7 1.000(0)

1FeTiO3 LiNbO3 phase TVFTHP2                              DISAGL   Version Win32
Mar 10 15:13:40 2009 Page  15

      Vector        Length       Optr Cell     Neighbor atom coordinates
       Fe1_Ti1     2.9624(26)    -1 0 0 1      0.00000  -0.13600   0.43580
       Fe1_O1      2.1093(28)     3 0 1 0      0.00000   0.23400   0.75000
       Fe1_O2      1.8787(18)     3 0 0 0      0.00000  -0.04800   0.61700
       Fe1_O2      2.0244(28)    -3 0 0 1      0.00000   0.04800   0.38300
       Fe1_O3      2.1854(20)    -3 0 0 1      0.00000   0.31100   0.43000
       Fe1_O3      1.9454(14)   103-1 0 0     -0.50000   0.18900   0.57000
       Fe1_O3      1.9454(14)   103 0 0 0      0.50000   0.18900   0.57000

      Angle         Degrees      atom 1 loc   atom 3 loc
       O1_Fe1_O2   100.55(8)      3 0 1 0      3 0 0 0
       O1_Fe1_O2   178.131(2)     3 0 1 0     -3 0 0 1
       O1_Fe1_O3    81.480(15)    3 0 1 0    103-1 0 0
       O1_Fe1_O3    81.480(15)    3 0 1 0    103 0 0 0
       O2_Fe1_O2    81.32(7)      3 0 0 0     -3 0 0 1
       O2_Fe1_O3   104.309(19)    3 0 0 0    103-1 0 0
       O2_Fe1_O3   104.309(19)    3 0 0 0    103 0 0 0
       O2_Fe1_O3    98.093(14)   -3 0 0 1    103-1 0 0
       O2_Fe1_O3    98.093(14)   -3 0 0 1    103 0 0 0
       O3_Fe1_O3   148.86(4)    103-1 0 0    103 0 0 0

      Vector        Length       Optr Cell     Neighbor atom coordinates

| Vector | Length | Optr Cell | Neighbor atom coordinates | | |
|---|---|---|---|---|---|
| Ti1_Fe1 | 2.9624(26) | -1 0 0 1 | 0.00000 | -0.13600 | 0.43580 |
| Ti1_Ti1 | 2.9624(26) | -1 0 0 1 | 0.00000 | -0.13600 | 0.43580 |
| Ti1_Ti1 | 3.1905(20) | -101-1 0 1 | -0.50000 | 0.36400 | 0.43580 |
| Ti1_Ti1 | 3.1905(20) | -101 0 0 1 | 0.50000 | 0.36400 | 0.43580 |
| Ti1_Ti2 | 3.219(5) | 1 0 0 0 | 0.00000 | 0.18900 | 0.25000 |
| Ti1_Ti2 | 3.1584(23) | 103-1 0 0 | -0.50000 | 0.31100 | 0.75000 |
| Ti1_Ti2 | 3.1584(23) | 103 0 0 0 | 0.50000 | 0.31100 | 0.75000 |
| Ti1_O1 | 2.1093(28) | 3 0 1 0 | 0.00000 | 0.23400 | 0.75000 |
| Ti1_O2 | 1.8787(18) | 3 0 0 0 | 0.00000 | -0.04800 | 0.61700 |
| Ti1_O2 | 2.0244(28) | -3 0 0 1 | 0.00000 | 0.04800 | 0.38300 |
| Ti1_O3 | 2.1854(20) | -3 0 0 1 | 0.00000 | 0.31100 | 0.43000 |
| Ti1_O3 | 1.9454(14) | 103-1 0 0 | -0.50000 | 0.18900 | 0.57000 |
| Ti1_O3 | 1.9454(14) | 103 0 0 0 | 0.50000 | 0.18900 | 0.57000 |

| Angle | Degrees | atom 1 loc | atom 3 loc |
|---|---|---|---|
| O1_Ti1_O2 | 100.55(8) | 3 0 1 0 | 3 0 0 0 |
| O1_Ti1_O2 | 178.131(2) | 3 0 1 0 | -3 0 0 1 |
| O1_Ti1_O3 | 101.32(10) | 3 0 1 0 | -3 0 0 1 |
| O1_Ti1_O3 | 81.480(15) | 3 0 1 0 | 103-1 0 0 |
| O1_Ti1_O3 | 81.480(15) | 3 0 1 0 | 103 0 0 0 |
| O2_Ti1_O2 | 81.32(7) | 3 0 0 0 | -3 0 0 1 |
| O2_Ti1_O3 | 158.129(24) | 3 0 0 0 | -3 0 0 1 |
| O2_Ti1_O3 | 104.309(19) | 3 0 0 0 | 103-1 0 0 |
| O2_Ti1_O3 | 104.309(19) | 3 0 0 0 | 103 0 0 0 |
| O2_Ti1_O3 | 76.81(10) | -3 0 0 1 | -3 0 0 1 |
| O2_Ti1_O3 | 98.093(14) | -3 0 0 1 | 103-1 0 0 |
| O2_Ti1_O3 | 98.093(14) | -3 0 0 1 | 103 0 0 0 |
| O3_Ti1_O3 | 79.025(22) | -3 0 0 1 | 103-1 0 0 |
| O3_Ti1_O3 | 79.025(22) | -3 0 0 1 | 103 0 0 0 |
| O3_Ti1_O3 | 148.86(4) | 103-1 0 0 | 103 0 0 0 |

| Vector | Length | Optr Cell | Neighbor atom coordinates | | |
|---|---|---|---|---|---|
| Fe2_O1 | 2.0199(14) | 101-1-1 0 | -0.50000 | 0.26600 | 0.25000 |
| Fe2_O1 | 2.0199(14) | 101 0-1 0 | 0.50000 | 0.26600 | 0.25000 |
| Fe2_O2 | 1.9270(19) | 1 0 0 0 | 0.00000 | 0.04800 | 0.11700 |

1FeTiO3 LiNbO3 phase TVFTHP2                    DISAGL   Version Win32
Mar 10 15:13:40 2009 Page  16

| Vector | Length | Optr Cell | Neighbor atom coordinates | | |
|---|---|---|---|---|---|
| Fe2_O2 | 1.9270(19) | -3 0 0 1 | 0.00000 | 0.04800 | 0.38300 |
| Fe2_O3 | 2.1768(26) | 1 0 0 0 | 0.00000 | 0.31100 | 0.07000 |
| Fe2_O3 | 2.1768(26) | -3 0 0 1 | 0.00000 | 0.31100 | 0.43000 |

| Angle | Degrees | atom 1 loc | atom 3 loc |
|---|---|---|---|
| O1_Fe2_O1 | 136.17(5) | 101-1-1 0 | 101 0-1 0 |
| O1_Fe2_O2 | 105.505(27) | 101 -1 -1 0 | 1 0 0 0 |
| O1_Fe2_O2 | 105.505(27) | 101 -1 -1 0 | -3 0 0 1 |
| O1_Fe2_O2 | 105.505(27) | 101 0 -1 0 | 1 0 0 0 |
| O1_Fe2_O2 | 105.505(27) | 101 0 -1 0 | -3 0 0 1 |
| O2_Fe2_O2 | 88.50(11) | 1 0 0 0 | -3 0 0 1 |

| Vector | Length | Optr Cell | Neighbor atom coordinates |
|---|---|---|---|
| Ti2_Ti1 | 3.219(5) | 1 0 0 0 | 0.00000  0.13600  0.56420 |
| Ti2_Ti1 | 3.219(5) | -3 0 0 1 | 0.00000  0.13600  -0.06420 |
| Ti2_Ti1 | 3.1584(23) | 103 -1 0 -1 | -0.50000  0.36400  0.06420 |
| Ti2_Ti1 | 3.1584(23) | 103 0 0 -1 | 0.50000  0.36400  0.06420 |
| Ti2_Ti1 | 3.1584(23) | -101 -1 0 1 | -0.50000  0.36400  0.43580 |
| Ti2_Ti1 | 3.1584(23) | -101 0 0 1 | 0.50000  0.36400  0.43580 |
| Ti2_O1 | 2.0199(14) | 101 -1 -1 0 | -0.50000  0.26600  0.25000 |
| Ti2_O1 | 2.0199(14) | 101 0 -1 0 | 0.50000  0.26600  0.25000 |
| Ti2_O2 | 1.9270(19) | 1 0 0 0 | 0.00000  0.04800  0.11700 |
| Ti2_O2 | 1.9270(19) | -3 0 0 1 | 0.00000  0.04800  0.38300 |
| Ti2_O3 | 2.1768(26) | 1 0 0 0 | 0.00000  0.31100  0.07000 |
| Ti2_O3 | 2.1768(26) | -3 0 0 1 | 0.00000  0.31100  0.43000 |

| Angle | Degrees | atom 1 loc | atom 3 loc |
|---|---|---|---|
| O1_Ti2_O1 | 136.17(5) | 101 -1 -1 0 | 101 0 -1 0 |
| O1_Ti2_O2 | 105.505(27) | 101 -1 -1 0 | 1 0 0 0 |
| O1_Ti2_O2 | 105.505(27) | 101 -1 -1 0 | -3 0 0 1 |
| O1_Ti2_O3 | 78.184(25) | 101 -1 -1 0 | 1 0 0 0 |
| O1_Ti2_O3 | 78.184(25) | 101 -1 -1 0 | -3 0 0 1 |
| O1_Ti2_O2 | 105.505(27) | 101 0 -1 0 | 1 0 0 0 |
| O1_Ti2_O2 | 105.505(27) | 101 0 -1 0 | -3 0 0 1 |
| O1_Ti2_O3 | 78.184(25) | 101 0 -1 0 | 1 0 0 0 |
| O1_Ti2_O3 | 78.184(25) | 101 0 -1 0 | -3 0 0 1 |
| O2_Ti2_O2 | 88.50(11) | 1 0 0 0 | -3 0 0 1 |
| O2_Ti2_O3 | 79.03(11) | 1 0 0 0 | 1 0 0 0 |
| O2_Ti2_O3 | 167.526(5) | 1 0 0 0 | -3 0 0 1 |
| O2_Ti2_O3 | 167.526(5) | -3 0 0 1 | 1 0 0 0 |
| O2_Ti2_O3 | 79.03(11) | -3 0 0 1 | -3 0 0 1 |
| O3_Ti2_O3 | 113.45(10) | 1 0 0 0 | -3 0 0 1 |

| Vector | Length | Optr Cell | Neighbor atom coordinates |
|---|---|---|---|
| O1_Fe1 | 2.1093(28) | 3 0 1 -1 | 0.00000  0.86400  0.06420 |
| O1_Fe1 | 2.1093(28) | -1 0 1 1 | 0.00000  0.86400  0.43580 |
| O1_Ti1 | 2.1093(28) | 3 0 1 -1 | 0.00000  0.86400  0.06420 |
| O1_Ti1 | 2.1093(28) | -1 0 1 1 | 0.00000  0.86400  0.43580 |
| O1_Fe2 | 2.0199(14) | 101 -1 0 0 | -0.50000  0.68900  0.25000 |
| O1_Fe2 | 2.0199(14) | 101 0 0 0 | 0.50000  0.68900  0.25000 |
| O1_Ti2 | 2.0199(14) | 101 -1 0 0 | -0.50000  0.68900  0.25000 |
| O1_Ti2 | 2.0199(14) | 101 0 0 0 | 0.50000  0.68900  0.25000 |

|   Angle     | Degrees    | atom 1 loc | atom 3 loc |
|-------------|------------|------------|------------|
| Fe1_O1_Fe1  | 125.89(9)  | 3 0 1 -1   | -1 0 1 1   |
| Fe1_O1_Ti1  | 0.000(0)   | 3 0 1 -1   | 3 0 1 -1   |

1FeTiO3 LiNbO3 phase TVFTHP2                    DISAGL   Version Win32
Mar 10 15:13:40 2009 Page  17

|   Angle     | Degrees     | atom 1 loc | atom 3 loc  |
|-------------|-------------|------------|-------------|
| Fe1_O1_Ti1  | 125.89(9)   | 3 0 1 -1   | -1 0 1 1    |
| Fe1_O1_Fe2  | 99.773(22)  | 3 0 1 -1   | 1 0 1 -1 0 0|
| Fe1_O1_Fe2  | 99.773(22)  | 3 0 1 -1   | 1 0 1 0 0 0 |
| Fe1_O1_Ti2  | 99.773(22)  | 3 0 1 -1   | 1 0 1 -1 0 0|
| Fe1_O1_Ti2  | 99.773(22)  | 3 0 1 -1   | 1 0 1 0 0 0 |
| Fe1_O1_Ti1  | 125.89(9)   | -1 0 1 1   | 3 0 1 -1    |
| Fe1_O1_Ti1  | 0.000(0)    | -1 0 1 1   | -1 0 1 1    |
| Fe1_O1_Fe2  | 99.773(22)  | -1 0 1 1   | 1 0 1 -1 0 0|
| Fe1_O1_Fe2  | 99.773(22)  | -1 0 1 1   | 1 0 1 0 0 0 |
| Fe1_O1_Ti2  | 99.773(22)  | -1 0 1 1   | 1 0 1 -1 0 0|
| Fe1_O1_Ti2  | 99.773(22)  | -1 0 1 1   | 1 0 1 0 0 0 |
| Ti1_O1_Ti1  | 125.89(9)   | 3 0 1 -1   | -1 0 1 1    |
| Ti1_O1_Fe2  | 99.773(22)  | 3 0 1 -1   | 1 0 1 -1 0 0|
| Ti1_O1_Fe2  | 99.773(22)  | 3 0 1 -1   | 1 0 1 0 0 0 |
| Ti1_O1_Ti2  | 99.773(22)  | 3 0 1 -1   | 1 0 1 -1 0 0|
| Ti1_O1_Ti2  | 99.773(22)  | 3 0 1 -1   | 1 0 1 0 0 0 |
| Ti1_O1_Fe2  | 99.773(22)  | -1 0 1 1   | 1 0 1 -1 0 0|
| Ti1_O1_Fe2  | 99.773(22)  | -1 0 1 1   | 1 0 1 0 0 0 |
| Ti1_O1_Ti2  | 99.773(22)  | -1 0 1 1   | 1 0 1 -1 0 0|
| Ti1_O1_Ti2  | 99.773(22)  | -1 0 1 1   | 1 0 1 0 0 0 |
| Fe2_O1_Fe2  | 136.17(5)   | 1 0 1 -1 0 0 | 1 0 1 0 0 0 |
| Fe2_O1_Ti2  | 0.000(0)    | 1 0 1 -1 0 0 | 1 0 1 -1 0 0|
| Fe2_O1_Ti2  | 136.17(5)   | 1 0 1 -1 0 0 | 1 0 1 0 0 0 |
| Fe2_O1_Ti2  | 136.17(5)   | 1 0 1 0 0 0 | 1 0 1 -1 0 0|
| Fe2_O1_Ti2  | 0.000(0)    | 1 0 1 0 0 0 | 1 0 1 0 0 0 |
| Ti2_O1_Ti2  | 136.17(5)   | 1 0 1 -1 0 0 | 1 0 1 0 0 0 |

| Vector  | Length     | Optr Cell | Neighbor atom coordinates      |
|---------|------------|-----------|--------------------------------|
| O2_Fe1  | 1.8787(18) | 3 0 0 -1  | 0.00000  -0.13600   0.06420    |
| O2_Fe1  | 2.0244(28) | -3 0 0 1  | 0.00000   0.13600  -0.06420    |
| O2_Ti1  | 1.8787(18) | 3 0 0 -1  | 0.00000  -0.13600   0.06420    |
| O2_Ti1  | 2.0244(28) | -3 0 0 1  | 0.00000   0.13600  -0.06420    |
| O2_Fe2  | 1.9270(19) | 1 0 0 0   | 0.00000   0.18900   0.25000    |
| O2_Ti2  | 1.9270(19) | 1 0 0 0   | 0.00000   0.18900   0.25000    |

|   Angle    | Degrees   | atom 1 loc | atom 3 loc |
|------------|-----------|------------|------------|
| Fe1_O2_Fe1 | 98.68(7)  | 3 0 0 -1   | -3 0 0 1   |

```
Fe1_O2_Ti1       0.000(0)      3 0 0-1    3 0 0-1
Fe1_O2_Ti1      98.68(7)       3 0 0-1   -3 0 0 1
Fe1_O2_Fe2     152.258(25)     3 0 0-1    1 0 0 0
Fe1_O2_Ti2     152.258(25)     3 0 0-1    1 0 0 0
Fe1_O2_Ti1      98.68(7)      -3 0 0 1    3 0 0-1
Fe1_O2_Ti1       0.000(0)     -3 0 0 1   -3 0 0 1
Fe1_O2_Fe2     109.06(10)     -3 0 0 1    1 0 0 0
Fe1_O2_Ti2     109.06(10)     -3 0 0 1    1 0 0 0
Ti1_O2_Ti1      98.68(7)       3 0 0-1   -3 0 0 1
Ti1_O2_Fe2     152.258(25)     3 0 0-1    1 0 0 0
Ti1_O2_Ti2     152.258(25)     3 0 0-1    1 0 0 0
Ti1_O2_Fe2     109.06(10)     -3 0 0 1    1 0 0 0
Ti1_O2_Ti2     109.06(10)     -3 0 0 1    1 0 0 0
Fe2_O2_Ti2       0.000(0)      1 0 0 0    1 0 0 0

 Vector      Length        Optr Cell     Neighbor atom coordinates
 O3_Fe1     2.1854(20)     -3 0 0 1      0.00000   0.13600  -0.06420
 O3_Fe1     1.9454(14)     103-1 0-1    -0.50000   0.36400   0.06420
 O3_Fe1     1.9454(14)     103 0 0-1     0.50000   0.36400   0.06420
```
1FeTiO3 LiNbO3 phase TVFTHP2                           DISAGL   Version Win32
Mar 10 15:13:40 2009 Page  18

```
 Vector      Length        Optr Cell     Neighbor atom coordinates
 O3_Ti1     2.1854(20)     -3 0 0 1      0.00000   0.13600  -0.06420
 O3_Ti1     1.9454(14)     103-1 0-1    -0.50000   0.36400   0.06420
 O3_Ti1     1.9454(14)     103 0 0-1     0.50000   0.36400   0.06420
 O3_Fe2     2.1768(26)      1 0 0 0      0.00000   0.18900   0.25000
 O3_Ti2     2.1768(26)      1 0 0 0      0.00000   0.18900   0.25000

   Angle        Degrees       atom 1 loc  atom 3 loc
 Fe1_O3_Fe1    148.86(4)      103-1 0-1   103 0 0-1
 Fe1_O3_Ti1    100.975(22)    103-1 0-1   -3 0 0 1
 Fe1_O3_Ti1      0.000(0)     103-1 0-1   103-1 0-1
 Fe1_O3_Ti1    148.86(4)      103-1 0-1   103 0 0-1
 Fe1_O3_Ti2     99.878(17)    103-1 0-1    1 0 0 0
 Fe1_O3_Ti1    100.975(22)    103 0 0-1   -3 0 0 1
 Fe1_O3_Ti1    148.86(4)      103 0 0-1   103-1 0-1
 Fe1_O3_Ti1      0.000(0)     103 0 0-1   103 0 0-1
 Fe1_O3_Ti2     99.878(17)    103 0 0-1    1 0 0 0
 Ti1_O3_Ti1    100.975(22)    -3 0 0 1    103-1 0-1
 Ti1_O3_Ti1    100.975(22)    -3 0 0 1    103 0 0-1
 Ti1_O3_Ti2     95.10(11)     -3 0 0 1     1 0 0 0
 Ti1_O3_Ti1    148.86(4)      103-1 0-1   103 0 0-1
 Ti1_O3_Ti2     99.878(17)    103-1 0-1    1 0 0 0
 Ti1_O3_Ti2     99.878(17)    103 0 0-1    1 0 0 0
```